\documentclass[aps,floatfix,notitlepage, tightenlines, nofootinbib]{revtex4-1}
\usepackage{amsmath, amssymb, amsfonts, amsthm, latexsym, epsfig, mathrsfs, xcolor, bbm, slashed}
\usepackage[top=1in, bottom=1in, left=.9in, right=.9in]{geometry}
\usepackage{mdframed}
\DeclareMathAlphabet{\mathpzc}{OT1}{pzc}{m}{it}
\usepackage[cal=boondox, calscaled=.94, bb=ams, scr=rsfs]{mathalfa} 
\usepackage[T1]{fontenc} % for thorn and eth
\usepackage{amsfonts}
\usepackage{graphicx}
\usepackage{amsmath}
\usepackage{color}

\usepackage[colorlinks=true, citecolor=blue,urlcolor=blue]{hyperref}
\usepackage{hyperref}
\usepackage{PeterStyle}
\usepackage{subfigure}
\usepackage{comment}
\usepackage{multirow}
\usepackage{calligra}

\newcommand{\Lie}{\pounds} 

\newcommand{\he}{\hat{e}}
\newcommand{\hmu}{\hat{\mu}}
\newcommand{\hA}{\hat{A}}

\newcommand{\hRp}{\hat{R}_{+}}
\newcommand{\hRm}{\hat{R}_{-}}
\newcommand{\hRpm}{\hat{R}_{\pm}}

\newcommand{\hRin}{\hat{R}_{\rm in}}

\newcommand{\hRmx}{\hat{R}_{\rm mix}}
\newcommand{\Rfin}{\tilde R_{\rm in}}
\newcommand{\Rfinf}{\tilde R_{\infty}}
\newcommand{\Rtfin}{\tilde{R}^{\rm far}_{\rm in}}
\newcommand{\Rtnin}{\tilde{R}^{\rm near}_{\rm in}}
\newcommand{\Rtfinf}{\tilde{R}^{\rm far}_{\infty}}
\newcommand{\Rtfp}{\tilde{R}^{\rm far}_{+}}
\newcommand{\Rtfm}{\tilde{R}^{\rm far}_{-}}
\newcommand{\h}{h}
\newcommand{\hp}{{h_+}}
\newcommand{\hm}{{h_-}}
\newcommand{\hpm}{{h_{\pm}}}
\newcommand{\G}[1]{\Gamma(#1)}
\newcommand{\NO}{{\mathcal N}}
\newcommand{\GO}{{\mathcal G}}
\newcommand{\rat}{{\mathcal{G}/\mathcal{N}}}
\newcommand{\ratf}{{\frac{{\mathcal{G}}}{\mathcal{N}}}}
\newcommand{\irat}{{\mathcal{N}/{\mathcal{G}}}}

\newcommand{\adst}{\mathrm{AdS}_2}
\newcommand{\SO}{SO}
\newcommand{\gff}{\tilde g^{\rm ff}}
\newcommand{\Gff}{\tilde G^{\rm ff}}
\newcommand{\gnf}{\tilde g^{\rm nf}}

\newcommand{\tr}{ {\tilde{r}} }
\newcommand{\tphi}{ {\tilde{\phi}} }

\begin{document}

\title{Scaling and Universality in Extremal Black Hole Perturbations}
\author{Samuel E.~Gralla\footnote{{\tt sgralla@email.arizona.edu}}}
\author{Peter~Zimmerman\footnote{{\tt peterzimmerman@email.arizona.edu}}}
\affiliation{Department of Physics, University of Arizona, \\
1118 E. Fourth Street, \\
Tucson, Arizona, 85721}
\date{\today}

\begin{abstract}
We show that the emergent near-horizon conformal symmetry of extremal black holes gives rise to universal behavior in perturbing fields, both near and far from the black hole horizon.  The scale-invariance of the near-horizon region entails power law time-dependence with three universal features: (1) the decay off the horizon is always precisely twice as fast as the decay on the horizon; (2) the special rates of $1/t$ off the horizon and $1/\sqrt{v}$ on the horizon commonly occur; and (3) sufficiently high-order transverse derivatives grow on the horizon (Aretakis instability).  The results are simply understood in terms of near-horizon ($\adst$) holography.  We first show how the general features follow from symmetry alone and then go on to present the detailed universal behavior of scalar, electromagnetic, and gravitational perturbations of $d$-dimensional electrovacuum black holes.
\end{abstract}
\maketitle
\newpage
\tableofcontents
\newpage
\section{Introduction and summary}\label{sec:intro}

In the greater enterprise of black hole physics, extremal black holes---those at the edge of the allowed parameter space---play a special role.  Their distinct mathematical properties generally demand separate analysis, while their privileged physical status gives them special interest.  
In astrophysical or condensed-matter applications, extremality corresponds to the interesting limits of high spin and low temperature, respectively.  In quantum gravity, the lack of Hawking radiation makes extremal black holes thermodynamically stable and hence easier to study.  The extremal limit has also seen a recent surge of interest from mathematicians interested in stability.

All of these various \textit{fora} for black holes---astrophysics, condensed matter physics, quantum gravity, and mathematics---involve in an essential way the study of their \textit{perturbations}.  One perturbs the spacetime and/or other fields involved, or for simplicity considers a test field propagating on the geometry.  From the behavior of these perturbing fields results many interesting quantities, such as the Hawking radiation spectrum, the propagator of a holographically dual theory, the gravitational-wave emission from some process, or the stability (or instability!) of the black hole itself.  While four-dimensional, asymptotically flat, electrovacuum black holes are remarkably constrained (they ``have no hair''), modern problems of interest increasingly demand higher and lower dimensions, additional fields like scalars and spinors, and/or the presence of vacuum energy that modifies the boundary behavior.  The complexity of this ``black hole zoo'' motivates the search for  \textit{universal} features, independent of the details of any specific case.

In physics quite generally, universal behavior emerges near special, ``critical'' points exhibiting emergent conformal symmetry.  In black hole physics, the near-horizon region of an extremal black hole functions as such a point, as it sees the emergence of the two-dimensional (global) conformal group $SO(2,1)$ as a spacetime symmetry \cite{Maldacena:1998uz,Bardeen:1999px,Kunduri:2007vf,Figueras:2008qh}.\footnote{For extensions to local conformal symmetries, see \cite{Guica:2008mu}.}  In this paper we find associated universality in perturbing fields both near \textit{and far} from the horizon.  In particular, we show that (under certain conditions) the near-horizon region gives rise to power law time-dependence\footnote{We assume a stationary black hole with $N$ commuting axisymmetries, and by ``time-dependence'' we mean Killing time along the orbits of a timelike linear combination of these Killing fields.  (The off-horizon decay rate does not depend on which Killing field one chooses; for an example see Eq.~\eqref{eq:an interesting thing} and discussion below.)  Here $t$ stands for any such notion of time, while $v$ is strictly the affine/Killing time along orbits of the degenerate horizon generators.  For scalars we refer to the value of the field, while for electromagnetic and gravitational perturbations we refer to a certain Hertz potential, from which the perturbation may be constructed.} in each angular mode, with three universal features: 
\begin{enumerate}
    \item The decay off the horizon is always precisely twice as fast as the decay on the horizon.
    \item The special rates of $1/t$ off the horizon and $1/\sqrt{v}$ on the horizon commonly occur \\ (i.e. over finite regions of parameter space, without fine tuning). 
    \item Sufficiently high-order transverse derivatives grow on the horizon (Aretakis instability). 
\end{enumerate}
If not swamped by other features (such as slower decay or an exponential instability), these rates will be visible at late times, as in the known $1/t$ \cite{Glampedakis:2001js} and $1/\sqrt{v}$ \cite{Casals:2016mel} tails of massless perturbations of extremal Kerr.  If not, they should still be identifiable at intermediate times, as in the transient $1/t$ decay occurring for charged perturbations of extremal Kerr-Newman \cite{Konoplya:2013rxa}.  At the very least, they can be identified with spectral analysis, as they are associated with a calculable special frequency in each example, often the superradiant bound.  The special rates of $1/t$ and $1/\sqrt{v}$ occur over large swaths of parameter space in our analysis, and we therefore expect that these rates will appear much more generally than the known examples, functioning as ``calling cards'' for an extremal black hole. 

The universality can be traced to the shared $\adst$ factor in extremal near-horizon geometries \cite{Kunduri:2007vf} and is simply understood in holographic terms.  Each angular mode features a special frequency near which the dynamics are governed by a field in $\adst$ with some scaling dimension $h$.  Elementary symmetry considerations force the bulk-boundary and boundary-boundary propagators to scale with exponents $h$ and $2h$, respectively,
\begin{align}\label{eq:ff}
    \mathcal{D} G_{\rm B \pd} = - h G_{\rm B\pd}, \qquad 
    \mathcal{D} G_{\rm \pd \pd} = -2 h G_{ \pd\pd},
\end{align}
where $\mathcal{D}$ is the action of an infinitesimal dilation.  The key observation is that these propagators encode the effects of $\adst$ on externally sourced perturbations (initial data of compact support away from the horizon), since the $\adst$ boundary functions as the gateway between near and far regions.  In particular, $G_{\pd \pd}$ governs fields that propagate in and out of the near-horizon region, corresponding to off-horizon properties of externally sourced fields, while $G_{\pd B}$ governs propagation only \textit{in} to the near-horizon region, corresponding to on-horizon properties.  The relative factor of $2$ in Eq.~\eqref{eq:ff} accounts for the first result above, while the bound $\textrm{Re}[h]\geq 1/2$ on $\adst$ scaling dimensions accounts for the second.  The third result, the Aretakis instability, is also a direct consequence of the symmetry  \eqref{eq:ff}  \cite{Gralla:2017lto}.  We flesh out these arguments in Sec.~\ref{sec:symmetry} below.

While the symmetry argument captures the essence of our results, it is far from the whole story.  In particular, the argument assumes that holographic propagators can be defined for some exponent $h$, which is possible only for certain choices of $\adst$ boundary conditions (typically Dirichlet).  In black hole perturbation problems, the $\adst$ boundary conditions are determined by the physics of the far region, and we are not free to adjust them to satisfy our holographic urges.  In fact, in many important cases, such as for Kerr black holes, these conditions are such that dynamics in pure $\adst$ would not even be well-posed!  
After presenting the symmetry argument in Sec.~\ref{sec:symmetry}, we go on to tell the full story in Secs.~\ref{sec:base space}-\ref{sec:full space} in terms of the range of boundary conditions that can arise in practice.  We delineate the parameter space where the basic results 1-3 survive, giving conditions that can be checked for any particular perturbation problem of interest.  We give an example of applying the formalism in Sec.~\ref{sec:examples}.

Some of the main features and results of our analysis have been noticed before in the holographic condensed matter literature.  In particular, beginning with Ref.~\cite{Faulkner:2009wj}, it was recognized that $\adst$ scaling behavior emerges at frequencies near the chemical potential, giving rise to power laws in the dual theory.  However, this body of literature has not, to our knowledge, considered decay on the horizon or discussed the growth of derivatives (Aretakis instability).  Another important difference is that the condensed matter literature focuses on asymptotically AdS black holes, for which $\adst$ instability (violation of the Breitenlohner-Freedman (BF) bound) typically implies a condensate-type (``superconducting'') instability of the spacetime.  For the asymptotically flat black holes included in our general framework, near-horizon BF-violation can instead entail the universal $1/t$ tail that we emphasize.  We may say that the $1/t$ gravitational-wave tail of extremal Kerr \cite{Glampedakis:2001js,Compere:2017hsi,Casals:2018eev} is an observational signature of $\adst$.  

The remainder of this paper is organized as follows.  In Sec.~\ref{sec:symmetry} we show how the $\mathfrak{so}(2,1)$ symmetries of $\adst$ with a uniform electric field dictate the main results.  In Sec.~\ref{sec:base space} we give a detailed study of charged, massive scalar fields in $\adst$ with a uniform electric field.  In Sec.~\ref{sec:SO(2,1) geometries} we study general near-horizon geometries, in which $\adst$ appears as a ``base space'' dictating the dynamics of each angular mode.  In Sec.~\ref{sec:full space} we discuss full extremal geometries.  In Sec.~\ref{sec:examples} we apply the formalism to perturbations of four dimensional Kerr-Newman-AdS.  Finally, App.~\ref{sec:planar RN-AdS} discusses non-compact horizons using the example of planar RN-AdS, while App.~\ref{sec:pain in the ax} discusses modes which require special care, which we call discrete.  Our notation is summarized in Tab.~\ref{tab:notationtable}.

\begin{table}[htbp]\caption{Table of Notation}
\centering % to have the caption near the table
\begin{tabular}{r  l}
\hline\hline
$ e$, $ \mu $ & \quad charge and mass of complex scalar field perturbation \\
$\hat e$, $\hat \mu $ & \quad charge and mass of $\adst$ perturbation \\
$ \hat P $ & \quad a quantity $P$ defined on the $\adst$ base space \\
$ \mathring P $ & \quad a quantity $P$ defined on the fiber space\\
$ \tilde P $ & \quad a quantity $P$ defined on the full geometry\\
$ \mu, \nu, \sigma, \rho, \ldots $ &\quad indices on the $d-2$-dimensional fiber space\\
$I,J,K, \ldots $ &\quad indices running over the $N$ azimuthal angles $\phi^I$ on the fiber space\\
$i,j,k,\ldots$ & \quad indices running over the $d-N-2$ coordinates $y^i$ on the fiber space\\
$a, b, c, d, \cdots $ & \quad indices for tensor fields on the $d$-dimensional near-horizon spacetime\\
$L , L'$ & \quad multi-index for multipoles
\end{tabular}
\label{tab:notationtable}
\end{table}

\section{Argument from symmetry}\label{sec:symmetry}

Before diving into the full calculations, we show how the main features can be derived from symmetry alone, subject to the presence of holography-friendly boundary conditions for perturbations of $\adst$.  In the full calculation, these conditions arise for the modes (called ``supplementary'' in our terminology) which respect the near-horizon BF bound.  The analysis of the BF-violating modes has many similar features, but does not lend itself as neatly to the language of holography, and we defer to the detailed calculations of Sec.~\ref{sec:base space}.

\subsection{Boundary and horizon coordinates and gauge}

The field equation for a charged field in $\adst$ accompanied by a uniform electric field arises in the near-zone dimensional reduction of the full perturbation problem near a special frequency.  The $\adst$ future Poincar\'e horizon is identified with the event horizon of the black hole, while the $\adst$ boundary is identified with an overlap region---colloquially the entrance to the throat region---where near and far expansions are matched.  It will be convenient to use separate coordinates and gauge when considering the horizon and the boundary.  In boundary-adapted coordinates and gauge, we have
\begin{align}\label{eq:ads2 stuff}
    \textrm{Boundary-adapted: \ } \qquad d\hat s^2=-x^2dt^2+\frac{dx^2}{x^2},\qquad \hA=xdt, 
\end{align}
which cover the Poincar\'e patch $x>0$.  (We use hats to distinguish $\adst$ quantities
---see Table ~\ref{tab:notationtable}.)  To discuss the horizon we use  ``ingoing'' coordinate and gauge
\begin{align}\label{eq:vAp}
    v=t-1/x, \qquad \hA'=\hA+d(\ln x) 
\end{align}
where
\begin{align}\label{eq:ads2-ingoing}
    \textrm{Horizon-adapted: \ } \qquad d\hat s^2 = -x^2 dv^2 +2 dv dx,\qquad \hA'=xdv.
\end{align}
The future horizon is described by $x=0$ in these coordinates.

\subsection{Symmetries}

A spacetime symmetry of a metric $g$ and electromagnetic gauge field $A$ is generated by a vector $X$ such that $\Lie_X g=0$ and $\Lie_X A = d f$ for some function $f$. The spacetime symmetry generators of \eqref{eq:ads2 stuff} are given by
\begin{align}\label{eq:generators}
 \hat{H}_0  = t\pd_t - x\pd_x, \qquad \hat{H}_+ = \pd_t, \qquad
 \hat{H}_- =  (t^2+1/x^{2}) \pd_t - 2 xt \pd_x,
\end{align}
which satisfy the $\mathfrak{so}(2,1)$ commutation relations $[ \hat H_+, \hat H_-] = 2 H_0$ and $[\hat H_{\pm},\hat H_0]= \mp \hat H_{\pm}$.  With our gauge choice, the gauge field is Lie-derived by $\hat H_0$ and $\hat H_+$ but not by $\hat H_-$.  In fact there is no gauge where $\hat A$ is invariant under all three generators.  This makes it convenient to introduce a generalized Lie derivative corresponding to a simultaneous infinitesimal change of coordinates and gauge \cite{Prabhu:2015vua, Zimmerman:2016qtn}.  We use an overbar to denote pairs of a vector field and a scalar,
\begin{align}
    \bar{X} = (X,\zeta),
\end{align}
and define the associated Lie derivative by the rules
\begin{subequations}\label{eq:Lie2}
\begin{align}
\bar{\Lie}_{\bar{X}} \hat{g} & = \Lie_X \hat{g} \\
\bar{\Lie}_{\bar{X}} \hat{A} & = \Lie_X \hat{A} +  d\zeta.
\end{align}
\end{subequations}
Charged fields will have similar derivative laws as appropriate.  For example, for a complex scalar $\hat\psi$ of charge $\hat{e}$ we define
\begin{align}\label{eq:Lie3}
    \bar{\Lie}_{\bar{X}} \hat{\psi} & = \Lie_X \hat{\psi} + i \hat{e} \zeta \hat{\psi}.
\end{align}
The commutator of two pairs acts only on the spacetime part: $ [\bar{X}_1,\bar{X}_2] := [X_1,X_2]$, where $\bar{X}_i = ( X_i, \zeta_i )$.
In this language the spacetime symmetries may be written in boundary-adapted coordinates and gauge as 
\begin{align}\label{eq:symm bnd}
    \textrm{Boundary-adapted: } \qquad \bar{H}_0 = ( \hat H_0, 0), \qquad \bar{H}_+ = ( \hat H_+, 0), \qquad \bar{H}_- = ( \hat H_-,-2/x).
\end{align}
These satisfy the $\mathfrak{so}(2,1)$ commutation relations while leaving the metric and gauge field invariant as $\Lie_{\bar X}\hat g =0$ and $\Lie_{\bar X} \hat A =0$.  Under a finite $U(1)$ gauge transformation $A \to A + d \Lambda$, the generator $\bar{X}=(X,\zeta)$ changes by $\zeta \to \zeta - \Lie_X \Lambda$.  Thus for ingoing coordinates and gauge \eqref{eq:vAp} we have (using $\Lambda= \ln x$)
\begin{align}\label{eq:symm hor}
    \textrm{Horizon-adapted: } \qquad \bar{H}_0 = ( \hat H_0,1), \qquad \bar{H}_+ = ( \hat H_+, 0), \qquad \bar{H}_- = ( \hat H_-, 2v ),
\end{align}
where the ingoing-coordinate representation of the Killing vectors is
\begin{align}\label{eq:ingoing H H H powah}
      \hat{H}_0 = v\pd_v - x\pd_x, \qquad \hat{H}_+ = \pd_v, \qquad \hat{H}_- = v^2\pd_v -2(xv+1) \pd_x.
\end{align}

\subsection{Power laws from symmetry}\label{eq:tails from symmetry}

Having introduced the relevant notion of symmetry, we now give a simple argument showing the existence of power law tails fixed by the scaling dimension set by the boundary asymptotics of  $\adst$ bulk fields.  The two-point function $G_{\rm BB}(v,x;v',x')$ of an $\adst$ field (BB for ``bulk-bulk'') must respect all the spacetime symmetries.  Respecting the dilation symmetry means
\begin{align}\label{eq:BB in da hizouse}
    \bar{\Lie}_{\bar{H}_0}G_{\rm BB} = 0,
\end{align}
where the generalized Lie derivative $\bar{\Lie}$ acts on both spacetime points of the two-point function.  In holography one defines a bulk-boundary propagator $G_{\rm B\pd}$ by selecting the exponent $h$ such that the following limit exists,
\begin{align}\label{eq:Bpd in da hizouse}
        G_{\rm B\pd}(t,x;t') & := \lim_{x' \to \infty} (x')^h G_{\rm BB}(t,x;t',x').
\end{align}
This limit is to be taken in boundary-adapted coordinates and gauge \eqref{eq:ads2 stuff}.  The symmetry \eqref{eq:BB in da hizouse} of the bulk-bulk propagator immediately implies
\begin{align}\label{eq:GBpd symm}
    \bar{\Lie}_{\bar{H}_0}G_{\rm B\pd} & = - h G_{\rm B\pd}.
\end{align}
That is, going to the boundary (or equivalently matching to the far-zone) breaks the full symmetry down to a scaling self-similarity.  
We may also take the bulk point to the boundary or to the horizon, defining boundary-boundary and horizon-boundary propagators by
\begin{align}\label{eq:1d in da hizouse}
    G_{\pd\pd}(t;t')  := \lim_{x \to \infty} x^h G_{\rm B \pd}(t,x;t'),
    \qquad 
    G_{\rm \mathcal{H} \pd}(v;t')  := \lim_{x \to 0} G_{\rm B\pd}(v,x;t').
\end{align}
For $ G_{\pd\pd}$ the limit is taken in boundary-adapted coordinates and gauge, while for $G_{\rm \mathcal{H} \pd}$ the limit is taken in horizon-adapted coordinates and gauge.\footnote{The horizon-boundary propagator relates points on the horizon to points on the boundary and hence appears naturally in ``mixed'' coordinates $v$ and $t'$.  Here $v$ is an affine parameter along a generator of the (degenerate) horizon, while $t$ is the time coordinate of the boundary theory.  Note, however, that we could equivalently use $v$ on the boundary, since $v=t-1/x \to t$ as $x \to \infty$.  To define $G_{\rm \mathcal{H} \pd}$ in a single coordinate system and gauge, we could take the boundary limit in the ingoing coordinates and gauge by suitably modifying the conformal factor $x^h \mapsto x^{h-i\hat{e}}$ in \eqref{eq:Bpd in da hizouse} to account for the gauge transformation.} The symmetry \eqref{eq:GBpd symm} [or \eqref{eq:BB in da hizouse}] now implies
\begin{align}\label{eq:symmmmmm}
    \bar{\Lie}_{\bar{H}_0}G_{\rm \pd\pd} = - 2 h G_{\rm \pd\pd}
    , \qquad \bar{\Lie}_{\bar{H}_0}G_{\rm \mathcal{H}\pd}  = -h G_{ \mathcal{H}\pd}.
\end{align}
That is, an additional $h$ appears for each point taken to the boundary.  The key observation now is that each of these objects is intrinsically one-dimensional, so the self-similarity only allows power laws.  Setting $t'=0$ without loss of generality,\footnote{The boundary-boundary and horizon-boundary propagators also inherit the time-translation symmetry $\bar{H}_+=(\pd_t,0)=(\pd_v,0)$ of the bulk-bulk propagator, meaning they can only depend on time differences $t-t'$ and $v-t'$, respectively.} from Eqs.~\eqref{eq:symm bnd} and \eqref{eq:symm hor} we see that $\bar{\Lie}_{\bar{H}_0}$ acts on $G_{\pd \pd}$ as $t \pd_t$ and on $G_{\mathcal{H}\pd}$ as $v \pd_v+i\he$.  Thus the precise power laws are 
\begin{align}
    G_{\rm \pd\pd} \propto t^{-2h}, \qquad G_{\rm \mathcal{H}\pd} \propto v^{-h-i\hat{e}}.
\end{align}
The boundary-boundary propagator governs perturbations that  propagate in and out of the near-horizon region, giving rise to a $t^{-2h}$ tail in the external region.  On the other hand, the horizon-boundary propagator governs perturbations that propagate only in, giving rise to a $v^{-h-i\hat{e}}$ tail on the horizon.  The decay is set by the real part of $h$,\footnote{Here we assume that the charge $\hat{e}$ is a real number.  In fact, for electromagnetic and gravitational perturbations the effective charge has an imaginary part.  However, the decay of invariants is still set by the real part of $h$---see discussion in Sec.~\ref{sec:EMGrav} below. \label{footnote:footsie}} which always differs by precisely a factor of 2 between the horizon and boundary correlators.  This is the first universal result mentioned in the introduction.

The second universal result requires a modicum of $\adst$ physics, which is the scaling dimension $h$ for Dirichlet boundary conditions of a charged scalar field,
\begin{align}\label{scaleit}
\h_+:= 1/2+\sqrt{1/4+\hmu^2-\he^2}.
\end{align}
The real part is at most $1/2$, giving rise to $1/t$ and $1/\sqrt{v}$ decay when this bound is saturated.  The scaling dimension is related to symmetry in that $h_+(h_+-1)$ is the Casimir of $\mathfrak{so}(2,1)$ on the boundary (Sec.~\ref{sec:scaling} below), but the precise formula \eqref{scaleit} requires the field equations.

\subsection{Aretakis instability from symmetry}\label{sec:Stefanos is so symmetric}

To see the Aretakis instability, we return to the self-similarity of the full bulk-boundary propagator \eqref{eq:symmmmmm}, generalizing arguments presented in \cite{Gralla:2017lto} in the context of the Kerr spacetime.  We again set $t'=0$ without loss of generality.  The equation can be solved in boundary- or horizon-adapated coordinates and gauge, where the generalized Lie derivative acts on $\adst$ scalars as 
\begin{subequations}
\begin{align}
    \textrm{Boundary-adapted: } \qquad \bar{\Lie}_{\bar{H}_0} & = t \pd_t - x \pd_x \label{eq:boundary H0}\\
    \textrm{Horizon-adapted: } \qquad \bar{\Lie}_{\bar{H}_0} & = v \pd_v - x \pd_x + i \hat{e}. \label{eq:horizon H0}
\end{align}
\end{subequations}
In the horizon-adapted gauge the general solution  to \eqref{eq:symmmmmm} is
\begin{align}\label{eq:general}
    G_{\rm B \pd} = v^{-h-i\hat{e}} f(xv),
\end{align}
for some function $f$.  By assumption the propagator is smooth on the horizon, so $f$ is smooth at $xv=0$.  It follows immediately that
\begin{align}\label{eq:Aretakis growth hormones}
    (\pd_x^n G_{\rm B \pd})|_{x=0} = v^{-h-i\hat{e} + n} f^{(n)}(0),
\end{align}
where $f^{(n)}$ is the $n^{\rm th}$ ordinary derivative.  That is, taking a transverse derivative adds a power of $v$, such that sufficiently high-order derivatives grow along the horizon---the Aretakis instability.   This implies that infalling observers experience large gradients \cite{Gralla:2016sxp}.  However, scalars constructed from the field remain small, since all such quantities will inherit self-similarity from \eqref{eq:GBpd symm} with some exponent $h'=nh$ where $n$ is a positive integer that counts the number of times the field appears in the formula for the scalar invariant.  The invariant then takes the general form \eqref{eq:general} with $h \to nh$, i.e. it decays on the horizon at the rate $v^{-n\textrm{Re}[h]}$.${}^{\ref{footnote:footsie}}$ Tensor fields and their decay can be treated in a similar way \cite{Gralla:2017lto}.  See also Refs.~\cite{Burko:2017eky,Hadar:2017ven} for complimentary discussions.

\subsection{Scaling dimension}\label{sec:scaling}

The precise scaling dimension $h$ requires the field equations and cannot follow from symmetry alone.  However, we can illustrate the role of the symmetries for charged scalars by noting that the wave operator can be written in terms of the generators as
\begin{align}\label{eq:groupie}
    \hat{D}^2 = \Omega + \hat{e}^2, \qquad \Omega := \bar\Lie_{\bar{H}_0} (\bar\Lie_{\bar{H}_0}- 1) - \bar\Lie_{\bar{H}_-} \bar\Lie_{\bar{H}_+}.
\end{align}
The quadratic operator $\Omega$ is the Casimir of $\mathfrak{so}(2,1)$.  In defining holographic propagators by limits involving multiplication by $x^h$, we have assumed boundary conditions such that the field goes as $x^{-h}$ at large $x$ in boundary-adapted coordinates and gauge.  The scaling self-similarity \eqref{eq:GBpd symm} of the bulk-boundary propagator means that $G_{\rm B\pd} \sim t^{-2h} x^{-h}$ at large $x$.  Using Eq.~\eqref{eq:groupie} one computes 
\begin{align}
    \Omega[t^{-2h} x^{-h}] = h(h-1)t^{-2h} x^{-h}\left( 1 + O(1/x^2) \right),
\end{align}
which is interpreted as ``$\Omega=h(h-1)$ on the boundary.''  In fact, the Casimir $\Omega$ is proportional to the identity on any irreducible representation, and $h(h-1)$ is a standard name.  Imposing the charged, massive scalar wave equation and using this relationship gives
\begin{align}\label{eq:ads2 covd to cas}
   0 & = \left( \hat{D}^2 - \hat{\mu}^2 \right) \hat{\psi} \\
     & = \left( \h(\h-1) + \hat{e}^2 - \hat{\mu}^2 \right) \hat{\psi} \nonumber.
\end{align}
The solutions are
\begin{align}\label{eq:hpm}
    \hpm=1/2 \pm \sqrt{1/4 + \hat{\mu}^2 - \hat{e}^2}.
\end{align}
These fix the large-$x$ behavior $x^{-\hpm}$ for bulk fields, as we recover explicitly in Sec.~\ref{sec:base space} below.

\section{Charged scalars in $\adst$ with a uniform electric field}\label{sec:base space}

\begin{table}
\begin{tabular}{|c|c|l|}
\hline
Name & Definition & \ \ \ \ \ \ \ \ \ \ \ \ \  Properties of $\hp$ \\
\hline
principal & $B<0, \ \textrm{not discrete}$ & $\hp \in \mathbb{C}, \quad \quad \hp = \tfrac{1}{2} + i r, \ \ \ r := \sqrt{|B|} $  \\
supplementary& $B>0, \ \textrm{not discrete}$ & $\hp \in \mathbb{R}^{>0}, \quad \hp = \tfrac{1}{2} + \sqrt{B} $\\
discrete & $h_++i\he \in \mathbb{Z}^{>0}$  &  \\
\hline
\end{tabular}
\caption{We categorize charged scalar fields in $\adst$ as principal, supplementary, or discrete depending on the values of the mass $\hat{\mu}$ and charge $\hat{e}$.  We have defined $\hp:=1/2+\sqrt{B}$ with $B:=1/4+\hat{\mu}^2-\hat{e}^2$.  This terminology originates in the  representation theory of the near-horizon symmetry group $SO(2,1)$ \cite{ClassicalAnalBrasu,Barut:1965}, although our definition of discrete differs when $\hat{e} \neq 0$.  The condition $B>0$ is also called the Breitenlohner-Freedman (BF) bound \cite{Breitenlohner:1982bm,Breitenlohner:1982jf,Ishibashi:2004wx, Kleban:2004bv,Holzegel:2011qj, Warnick:2012fi}. Properties of $h_-$ follow from the properties of $h_+$ using the relation $h_+ + h_-=1$.}
\label{tab:h}
\end{table}

We now give a detailed study of charged, massive scalar fields $\hat\psi$ in $\adst$ with a uniform electric field.  The field equation is
\begin{align}\label{eq:AdS2 charged scalar eq}
\left(\hat{D}^2-\hat{\mu}^2\right)\hat\psi=0, \qquad \hat{D}:=\hat{\nabla}-i\he \hat{A},
\end{align}
where $\hat{\nabla}$ is the metric-compatible derivative on $\adst$ and $\hat e$ is the scalar charge of the field.\footnote{Here we allow the charge $\hat{e}$ to be any complex number.  In the applications to full geometries that we consider below, $\hat{e}$ will be real for scalar perturbations and have integer imaginary part for electromagnetic and gravitational perturbations---see Eqs.~\eqref{eq:psi ads2 eq} and \eqref{eq:psib eq} below.}  We work in boundary-adapted coordinates and gauge \eqref{eq:ads2 stuff}.

The equation separates under the ansatz
\begin{equation}\label{eq:psi mode}
     \hat\psi=e^{-i\omega t}R(x),
\end{equation}
giving the radial equation
\begin{equation}\label{eq:rad eq ads2}
    \left[\frac{d}{dx}\left(x^2\frac{d}{dx}\right)+\frac{(\omega+\he x)^2}{x^2}-\hmu^2\right]R=0.
\end{equation}
A convenient set of solutions is represented as Whittaker functions \cite{NIST:DLMF}, which we denote by
\begin{align}\label{eq:Ads2 radial sols}
    \hRin=W_{i\he,\hp-1/2}(-2i\omega/x), \qquad  
    \hRpm=(-2i\omega)^{-\hpm} M_{i\he,\hpm-1/2}(-2i\omega/x),
\end{align}
where $\hpm$ was given previously in Eq.~\eqref{eq:hpm}.  Only two solutions are needed to span the general solution to \eqref{eq:rad eq ads2}, but we find it convenient to introduce all three.  Note, however, that when $\hp=1/2$ the $\hat{R}_\pm$ solutions are the same, and when $h_+=1,3/2,2,\dots$ the $\hat{R}_-$ solution does not exist.  Henceforth we will assume that $2h_+$ is not a positive integer (or equivalently that $2 h_-$ is not zero or a negative integer).  However, all expressions that are well-defined in the limit that $2 h_+$ becomes an integer do in fact hold in that case.

We refer to $\hpm$ as the weight of the bulk field or the scaling dimension of the corresponding CFT operator (see discussion in Sec.~\ref{sec:Diri/Neum conds} below).  The character of the weights (real, complex, or integer) determines many important features of the dynamics.  We summarize the important cases for our work in Table \ref{tab:h}. 

Generic solutions behave as a linear combination of $x^{-\hp}$ and $x^{-\hm}$ at the boundary $x \to \infty$.  The $R_\pm$ solutions are distinguished by behaving solely as $x^{-\hpm}$.  Their asymptotics are
\begin{align}\label{eq: Rplus asymptotics}
    \hRpm & \sim (-2i\omega)^{-\hpm} \begin{cases}\displaystyle
   \frac{\Gamma(2\hpm)}{\Gamma(\hpm-i\he)} e^{-i\omega/x}(-2i\omega/x)^{-i\he} + \frac{\Gamma(2\hpm)}{\Gamma(\hpm+i\he)}e^{i\omega/x + \epsilon(\hpm-i\he)\pi i} (-2i\omega/x)^{i \he}, \qquad &x \to 0, \\
    (-2i\omega/x)^{\hpm}, \qquad &x \to \infty, \end{cases}
\end{align}
where $\epsilon = \mathrm{sgn}\left(\mathrm{Im}(-i\omega)\right)$ assuming $x>0$. Here we have used Eq.~(7) in Sec.~6.7 of Ref.~\cite{Bateman:100233} Vol.~1.   Here, $\sim$ denotes asymptotic equality in the sense that $a(x) \sim b(x)$ as $x\to x_0$ is equivalent to $a(x)/b(x) \to 1$ as $x\to x_0$.  
Notice that each of the $R_{\pm}$ solutions has waves moving in both directions $e^{\pm i \omega/x}$ at the horizon $x \to 0$.  The $R_{\rm in}$ solution is distinguished by having only waves traveling into the black hole.  Its asymptotics are  
\begin{align}\label{eq:ads2 in asy}
     \hRin&\sim \begin{cases}\displaystyle
    e^{i\omega/x} (-2i\omega/x)^{i \hat e}, \qquad &x \to 0, \\
    A_+ x^{-\hp} + A_- x^{-\hm}, \qquad &x \to \infty,\end{cases}
\end{align}
where
\begin{equation}\label{eq:Apm}
    A_\pm=\frac{(-2i\hat\omega)^{\hpm}\Gamma(1-2\hpm)}{\Gamma(1-\hpm-i\he)}. 
\end{equation}

We will find it convenient to consider Green functions, which satisfy
\begin{align}
\hat D^2 \hat G = \delta_2,
\end{align}
where $\delta_2 = \delta(t-t')\delta(x-x')$ is the invariant delta-function of $\adst$.  We decompose  $\hat G$ into  frequencies by 
\begin{equation}\label{eq:hat G modes}
    \hat G = \frac{1}{2\pi}\int_{-\infty+ic}^{\infty+ic} e^{-i\omega(t-t')} \hat g(x,x')\, d \omega,
\end{equation}
where $c$ is a real number chosen to put the integration contour in the region where $\hat g$ is analytic.  (This is just the inverse Laplace transform with $s=-i\omega$, which agrees with the Fourier transform for causal propagation.\footnote{The inverse Laplace transform does not extend before $t=t'$, but for causal propagation $\hat{G}$ will vanish for these times anyway.  When we perform inverse Laplace transforms, below, we will regard $\hat{G}$ as vanishing for $t<t'$.  Our preference for Laplace over Fourier arises from the convenience of the former for initial value problems.})
In the mode decomposition for $\hat G$ we have introduced an $\adst$ ``transfer function'' $\hat g(x,x')$, which satisfies Eq.~\eqref{eq:rad eq ads2} with $\delta(x-x')$ on the right-hand side instead of zero,
\begin{align}\label{eq:Greenie}
    \left[\frac{d}{dx}\left(x^2\frac{d}{dx}\right)+\frac{(\omega+\he x)^2}{x^2}-\hmu^2\right]\hat g(x,x')=\delta(x-x').
\end{align}

\subsection{Dirichlet/Neumann conditions and holography}\label{sec:Diri/Neum conds}
Specifying the dynamics requires a choice of boundary conditions.  We will do so by making a choice of Green function $\hat{G}$.  The most common choices are to solve Eq.~\eqref{eq:Greenie} by
\begin{equation}\label{eq:g2}
    \hat g^{\pm}(x,x')=\frac{\hRin(x_<)\hRpm(x_>)}{\mathcal W_{\pm}},
\end{equation}
where $x_<$ and $x_>$ are (respectively) the lesser and greater of the points $x$ and $x'$ and $\mathcal W_{\pm} = x^2(\hRin\hRpm' - \hRin' \hRpm)$ is a constant given by (12.14.26 of Ref.~\cite{NIST:DLMF})\footnote{If $\hpm-i \hat{e} = 0$ then $\hat{R}_\pm$ becomes proportional to $\hat{R}_{\rm in}$ and all mode solutions satisfy both boundary conditions.  We assume $\hpm+i\hat{e} \neq 0$ to avoid this physical pathology.\label{footnote:hi}}
 \begin{equation}\label{eq:wron in pm}
     \mathcal  W_{\pm} = (-2i\omega)^{1-\hpm}\Gamma(2\hpm)/\Gamma(\hpm-i\he).
 \end{equation}
The presence of $\hat{R}_{\rm in}$ enforces causal propagation.  The use of $\hat g^+$ forces boundary behavior of $x^{-\hp}$, normally called ``Dirichlet'' conditions, while the use of $\hat g^-$ forces boundary behavior of $x^{-\hm}$ normally called ``Neumann'' conditions \cite{Warnick:2012fi}.\footnote{The Neumann condition may only be imposed when $1/2<h_+<1$ \ ($0<\hm>1/2$).  The Neumann version of all formulae below should be ignored when this is not satisfied.% \PZ{Lower bound is a unitary bound (complex h). Upper bound is a decay requirement at large $x$.}.
Note that in some references  $\hat R_+$ is called the Neumann mode because it is the one whose coefficient would be specified in Neumann boundary conditions.} 

These conditions allow the definition of a boundary field (``dual operator'') $\hat{\mathcal{O}}_{b,\pm}(t)$ by
\begin{align}\label{eq:boundary field}
    \hat{\mathcal{O}}_{b,\pm} = \lim_{x \to \infty} x^{\hpm} \hat\psi,
\end{align}
which can be regarded as living on the boundary $x \to \infty$ with metric $\lim_{x \to \infty}x^{-2} ds^2=-dt^2$.  The dilation symmetry $(x \to \lambda^{-1} x,t \to \lambda t)$ of the bulk induces a global conformal transformation $-dt^2 \to \lambda^{2} (-dt^2)$ on the boundary, under which the operator $\hat{\mathcal{O}}_{b,\pm}$ transforms as $\hat{\mathcal{O}}_{b,\pm} \to\lambda^{\h_{\pm}}\hat{\mathcal{O}}_{b,\pm}$.  In this sense $\h_{\pm}$ is the conformal scaling dimension of the dual operator.

The two-point function for the dual field is inherited from the bulk dynamics at large $x$.  We first define a bulk-boundary transfer function by taking one point to infinity and peeling off the leading behavior,
\begin{equation}\label{eq:g B pf def}
   \hat g^\pm_{\mathrm{B}\pd}:=\lim_{x'\to \infty} (x')^{\hpm}\hat g^\pm(x,x') = \frac{\Gamma(\hpm-i\he)}{\Gamma(2\hpm)}(-2i\omega)^{\hpm-1} \hRin(x).
\end{equation}
In this limit the inverse transform \eqref{eq:hat G modes} may be computed exactly from Eq.~(9) in Sec.~5.20 of Ref.~\cite{bateman1954tables}.\footnote{The original Bateman manuscript~\cite{Bateman:100233} gives a restricted range $\mathrm{Re}(h)<1$ for this transform.  However, by using the integral representation of the Whittaker $W$ function given in Eq.~(3.5.16) of \cite{slater1960confluent}, we have found that integral transform still applies when $\hp \pm i\he \neq 1,2,3,\ldots$.  In the present context this condition is equivalent to $\hp + i\he \neq 1,2,3,\ldots$, since $\hp - i \he = 1,2,3,\dots$ cannot occur [see Eq.~\eqref{scaleit}].} 
Provided that 
\begin{align}\label{eq:deg cas}
\hp+i\he \notin \mathbb{Z}^{>0},
\end{align}
the time-domain bulk-boundary Green function $\hat G^\pm_{\mathrm B\pd}$ \eqref{eq:Bpd in da hizouse} is given by
\begin{align}\label{eq:hat Gpm Bpd}%p7e17
\hat G^\pm_{\mathrm{B}\pd}&=C_\pm x^{h_\pm}\left(\frac{(tx-1)}{2}\right)^{-i\hat e-\hpm}\left(1+\frac{(tx-1)}{2}\right)^{i\hat e-\hpm}\Theta(tx-1),
\end{align}
with
\begin{align}\label{eq:Cpm}
    C_\pm := \frac{\Gamma(\hpm-i \hat{e})}{2\Gamma(2\hpm) \Gamma(1-\hpm - i \hat{e})}.
\end{align}
In Eq.~\eqref{eq:hat Gpm Bpd} we have set $t'=0$ without loss of generality; this quantity may be restored by sending $t \to t-t'$.  Notice that the $\Theta$ function enforces causality, since $t=1/x$ is the (one-sided) light cone of the boundary point $t'=0$.  The coefficient $C_\pm$  vanishes in the special case $h_+ + i \hat{e} \in \mathbb{Z}^{>0}$ that was excluded in Eq.~\eqref{eq:deg cas}; the meaning of this case in the full geometry is discussed in App.~\ref{sec:pain in the ax}.
Notice, however, that the bulk-boundary propagator does remain well-defined in the case that $\hp$ is a half-integer that was excluded earlier.  We have therefore computed the result for half-integer $\hp$ by analytic continuation; a direct computation is also possible.\footnote{ \label{Whitlog} If $\hp$ is a half-integer, the character of the Whittaker $W$ function changes by picking up a logarithmic dependence (see for example DLMF Eq.~(13.14.8) \cite{NIST:DLMF}). The inverse Laplace transform in this case may be performed using Eq.(6) in \cite{Casals:2016mel}. }

We may now compute the boundary-boundary correlator \eqref{eq:1d in da hizouse} by taking $x$ to infinity in Eq.~\eqref{eq:hat Gpm Bpd},
\begin{equation}\label{eq:Gpdpd}
     \hat G^\pm_{\pd\pd}:= \lim_{x\to\infty} x^{\hpm}\hat G^\pm_{B\pd} =
    C_\pm (t/2)^{-2\hpm}\Theta(t).
\end{equation}
In the holographic interpretation, this defines the dynamics of the dual operator.  Eq.~\eqref{eq:Gpdpd} also follows from the Son/Starinets \cite{Son:2002sd} prescription, where the frequency-domain boundary-boundary correlator is defined by ratios of $A_+$ and $A_-$ as
\begin{equation}\label{eq:G SS}
    \hat g_{\pd \pd}^\pm(\omega) := N_\pm \frac{A_\pm}{A_\mp}= \frac{\Gamma(1-2\hpm)\Gamma(\hpm-i\he)}{\Gamma(2\hpm)\Gamma(1-\hpm-i\he)}(-2i\omega)^{2\hpm-1}.
\end{equation}
where we have fixed the normalization $N_{\pm}=1/(2\hpm-1)$ to obtain exact agreement with Eq.~\eqref{eq:Gpdpd} after inverse Laplace (or Fourier) transform.  Below we will find use for an abbreviated notation, 
\begin{equation}\label{eq:barred quantities}
    \bar A_\pm = \mp \frac{\G{2-2\hpm}}{\G{1-\hpm-i\he}}, \qquad \GO := \frac{\bar A_+}{\bar A_-} = (2h_+-1)\frac{\Gamma(1-2\hp)\Gamma(\hp-i\he)}{\Gamma(2\hp)\Gamma(1-\hp-i\he)},
\end{equation}
in which the Dirichlet two-point function is
\begin{align}\label{eq:g pd pd script G}
 \hat g_{\pd \pd}^+(\omega) = \frac{\mathcal{G}}{2h_+-1} (-2i\omega)^{2\hp-1}.
\end{align}

The expression \eqref{eq:hat Gpm Bpd} for the bulk-boundary propagator is not regular on the horizon.  Changing the bulk point to horizon-regular coordinates and gauge \eqref{eq:ads2-ingoing} gives
\begin{align}\label{eq:hat Gpm Bpd in time}%p8e19
  \textrm{Horizon-adapted gauge: \ } \qquad \hat G^\pm_{\mathrm{B}\pd}= C_\pm \left(\frac{v}{2}\right)^{-i\hat e-\hpm}\left(1+\frac{vx}{2}\right)^{i\hat e-\hpm}\Theta(v). 
\end{align}
Again, causality is manifest as $v=0$ is the one-sided light cone of the boundary point $t'=0$.  Finally we may take $x \to 0$ to produce the boundary-horizon propagator \eqref{eq:1d in da hizouse},
\begin{align}\label{eq:GHpd}
      \hat G^\pm_{\mathcal{H}\pd}=C_\pm \left(\frac{v}{2}\right)^{-i\hat e-\hpm}\Theta(v). 
\end{align}
For effect, we now collect the results \eqref{eq:Gpdpd} and \eqref{eq:GHpd} 
and restore the $t'$ coordinate,
\begin{subequations}
\begin{align}
    \hat G^\pm_{\pd \pd}(t,t') & =  C_\pm \left(\tfrac{1}{2}(t-t')\right)^{-2h_\pm} \Theta(t-t'), \\
    \hat G^\pm_{\mathcal{H}\pd}(v,t') & = C_\pm \left(\tfrac{1}{2}(v-t')\right)^{-h_\pm - i\hat{e}} \Theta(v-t'),
\end{align}
\end{subequations}
where $C_\pm$ is given by Eq.~\eqref{eq:Cpm}.
These take the general form required by conformal symmetry \eqref{eq:symmmmmm} and causality, with the coefficient $C_\pm$ and the exponent $h_+$ arising from the details of $\adst$.

\subsection{Mixed boundary conditions}\label{sec:mixed bcs}

We now consider a more general scenario where the boundary condition set at infinity is a mixture of Neumann and Dirichlet, while preserving the ingoing boundary condition at $x=0$. For this, we introduce a new function $\hRmx$ as the linear combination
\begin{subequations}
\begin{align}\label{eq:sf R}
    \hRmx &:= B_+\hRp + B_-\hRm,\\
    &\sim B_+\,x^{-\hp} + B_- \,x^{-\hm},\qquad x\to\infty, \label{eq:sf R asy}
\end{align}
\end{subequations}
where $B_\pm$ may in general depend on $\omega$ (but not $x$).  The physics is contained in the ratio of $B_+$ and $B_-$, which we denote by $\mathcal{N}$,
\begin{align}\label{eq:N}
     \mathcal N:=\frac{B_+}{B_-}.
\end{align}
We are mainly interested in the case where $\mathcal{N}$ is independent of $\omega$, but for this section we leave it arbitrary.  The transfer function for perturbations respecting this ratio of falloffs is
\begin{equation}
\hat g^{\rm mix} = \frac{\hat R_{\rm in}(x_<)\hat R_{\rm mix}(x_>)}{\mathcal W_{\rm mix}},
\end{equation}
where $\mathcal{W}_{\rm mix}$ is a constant given by $x{'}^2$ times the Wronskian of the in and mix solutions.  For mixed boundary conditions, we assume that $2\hp$ is not an integer.  In this case we may use Eq.~(13.14.33) of \cite{NIST:DLMF} to express $\hat{R}_{\rm in}$ in terms of the barred coefficients introduced in \eqref{eq:barred quantities},
\begin{equation}\label{eq:Rin from Rp Rm}
    \hRin = \bar{A}_- \frac{\GO(-2i\omega)^{\hp}\hRp+ (-2i\omega)^{\hm}\hRm}{2\hp-1}.
\end{equation}
The mixed transfer function may then be written
\begin{equation}\label{eq:sf hat g NR}%p11e23 
    \hat{g}_{\rm mix} = \frac{\left((-2i\omega)^{2\hp-1}\GO\hRp(x_<)+\hRm(x_<)\right)\left(\mathcal N \hRp(x_>)+\hRm(x_>)\right)}{(1-2\hp)\left(\mathcal N-\GO(-2i\omega)^{2\hp-1}\right)},
\end{equation}
where again $x_<$ and $x_>$ are respectively the lesser and greater of $x$ and $x'$.
Observe that with these mixed boundary conditions, the denominator (i.e. $\mathcal W$) is now a difference of terms.  Consequently, the analytic structure of the  transfer function has changed nontrivially, with new poles and additional branch structure having emerged.

For Dirichlet or Neumann conditions, we proceeded to define bulk-boundary and boundary-boundary correlators by peeling off leading behavior at large values of $x$.  In the mixed case this is not possible because both behaviors $x^{-\hpm}$ appear.  However, we will still need the large-$x$ asymptotics, which [using \eqref{eq:sf R asy}] are given by 
\begin{align}\label{eq:hat sf g xp large}%p12e25
\hat{g}_{\rm mix}(x,x'\to\infty) &\sim\frac{\left((-2i\omega)^{2\hp-1}\GO\hRp+\hRm\right)\left(\mathcal Nx'{}^{-\hp}+x'{}^{-\hm}\right)}{(1-2\hp)\left(\mathcal N-\GO(-2i\omega)^{2\hp-1}\right)},
\end{align}
and
\begin{align}\label{eq:hat sf g x xp large}%p12e25
\hat{g}_{\rm mix}(x\to\infty,x'\to\infty) &\sim\frac{\left((-2i\omega)^{2\hp-1}\GO x^{-\hp}+x^{-\hm}\right)\left(\mathcal Nx'{}^{-\hp}+x'{}^{-\hm}\right)}{(1-2\hp)\left(\mathcal N-\GO(-2i\omega)^{2\hp-1}\right)}.
\end{align}

\subsection{Frequency-independent mixed boundary conditions}\label{sec:mixed special BCs}

The mixed boundary conditions are characterized by prescribing the ratio $\mathcal{N}(\omega)$ of the two boundary behaviors.  We are mainly interested in the case where this ratio is independent of $\omega$, and we henceforth assume
\begin{align}
    \pd_\omega \mathcal N = 0.
\end{align}
The important quantity to consider is
\begin{align}\label{eq:chi}
    \chi := \mathcal{N} - \mathcal{G}(-2 i \omega)^{2\hp-1},
\end{align}
which appears in the demoninator of the transfer function \eqref{eq:sf hat g NR} (and \eqref{eq:hat sf g xp large} and \eqref{eq:hat sf g x xp large}).  At this stage it is useful to treat the principal and supplementary cases (Tab.~\ref{tab:h}) separately.  For principal fields where $\hp=1/2+i r$ with $r>0$, we need three further sub-cases:
\begin{subequations}\label{eq:principal cases}
\begin{align}
    \textrm{principal case I:  \ \ \ \ \ \ \ \ \ \ \ } \qquad & |\mathcal{G}/\mathcal{N}| < e^{-\pi r}\label{eq:prince case I} \\
    \textrm{principal case II:  }\qquad  e^{-\pi r} \leq &  |\mathcal{G}/\mathcal{N}| \leq e^{3 \pi r}\label{eq:prince case II} \\
    \textrm{principal case III: \ \ \ \ \ \ \ \ \ \ } \qquad & |\mathcal{G}/\mathcal{N}| > e^{3 \pi r} \label{eq:prince case III}.
\end{align}
\end{subequations}
The supplementary case further requires a fourth independent analysis.  We now give each in turn.

\subsubsection{Principal case I}\label{eq:mixed principal I}

In case I, the first term in $\chi$ dominates (its modulus is always larger than that of the second term), and we may expand $1/\chi$ in a geometric series in $(-2 i \omega)^{2\hp-1} \mathcal{G}/\mathcal{N}$.  This allows us to invert the large-$x'$ transfer function term-by-term via Eq.~(9) in Sec.5.20 of Ref.~\cite{bateman1954tables},\footnote{As before, the integral transform does not hold in the case $h_+ + i \hat{e} \in \mathbb{Z}^{>0}$.  However, this special case cannot arise for a principal mode for which the imaginary part of the charge is not half-integer, so there is no corresponding restriction on \eqref{eq:sf ginvP1}.} giving
\begin{align}\label{eq:sf ginvP1}
  &\hat{G}_{\rm mix}(t,x,x'\to\infty) \sim\frac{1}{\bar{A}_-\mathcal N}\left(\mathcal N x'{}^{-\hp}+x'{}^{-\hm}\right)x^{-i\he}\left(\frac{t-1/x}{2}\right)^{-1/2-ir-i\he}\Theta(t-1/x)\\
   &\times\sum_{n=0}^\infty\left(\ratf\right)^n\left(\frac{t-1/x}{2}\right)^{-2irn} {}_2\tilde{F}_1\left(\hp-i\he,1-\hp-i\he;1-\hp-i\he+(1-2\hp)n;-(tx-1)/2\right), \nonumber
\end{align}
where ${}_2 \tilde F_1(a,b;c;z) = {}_2F_1(a,b;c;z)/\Gamma(c)$.
Applying Eq.~(15.8.2) of \cite{NIST:DLMF}, we find that the large-$x$ behavior of \eqref{eq:sf ginvP1} is
\begin{align}\label{eq:G xx' big 1}
   \hat{G}_{\rm mix}(t,x\to\infty,x'\to\infty) \sim  & \left(\frac{2}{t}\right)^{1+ir} \frac{\left(\mathcal N x'{}^{-ir}+x'{}^{ir}\right)}{2ir\mathcal N \sqrt{xx'}}\Theta(t) \sum_{n=0}^\infty\left(\ratf\right)^n\left(\frac{2}{t}\right)^{2irn}
  \left( \frac{(tx/2)^{ir}}{\G{-2irn}}+ \frac{\mathcal G(tx/2)^{-ir}}{\G{-2ir(n+1)}}\right)
\end{align}
after substituting for $\hpm=1/2\pm ir$.  In the limit $\mathcal{N} \to \infty$ we recover the Dirichlet result $(x')^{-h_+}\hat{G}^+_{\mathrm{B} \pd}$ upon restoring $h_+$ in favor of $r$.

\subsubsection{Principal case II}

In case II, the two terms in $\chi$ are the same order of magnitude, and there are infinitely many zeros, corresponding to quasinormal mode frequencies $\omega_n$.  We are unable to invert the Laplace transform exactly in this case, but we can give the spectrum of quasinormal modes as\footnote{To demonstrate this claim let $z=-2i\omega$. Then the pole condition reads $z^{-2ir}=\rat$, or $e^{2r\arg z}e^{-2ir\ln\abs{z}}=\abs{\rat}e^{i\arg\rat}$. It follows that $\arg{z}=\frac{1}{2r}\ln\abs{\rat}$ and $\abs{z} = \exp \left[- \frac{1}{2r}\left(\arg(\rat) + 2 \pi n \right)\right]$. Now, as $\omega = \frac12 i z$, one finds  $\omega = \frac{i}{2}\abs{z}\left(\cos\arg \,z + i \sin\arg \, z \right)$.}
\begin{equation}\label{eq:spec prince}
\displaystyle
\omega_n=-\frac{1}{2}e^{-\frac{1}{2r}\left(\arg(\rat)+2\pi n\right)}
    \Bigg[\sin\left(\frac{\ln\abs{\rat}}{2r}\right)-i\cos\left(\frac{\ln\abs{\rat}}{2r}\right)\Bigg],\qquad n\in\mathbb Z.
\end{equation}
Upon inspecting the imaginary part of \eqref{eq:spec prince}, we see that a mode instability occurs when 
\begin{equation}\label{eq:QNM instability crit prince}
 e^{-\pi r}<\abs{\rat}<e^{\pi r}, \quad \text{(instability criterion, principal case)}.
\end{equation}

\subsubsection{Principal case III}

In case III, the second term in $\chi$ dominates and we may expand in $\mathcal{N}/(\mathcal{G}(-2 i \omega)^{2\hp-1}))$.  This turns out to be simply related to case I by
\begin{align}
    \hat{G}_{\rm mix}(t,x,x'\to\infty) \sim \textrm{RHS of \eqref{eq:sf ginvP1} with $h_+ \to h_-$ and $\mathcal{N} \to 1/\mathcal{N}$.}
\end{align}
(The same substitution can also be made in \eqref{eq:G xx' big 1} to obtain the large-$x$ limit.)  Note that under $h_+ \to h_-$, we have $r \to -r$ and $\mathcal{G} \to 1/\mathcal{G}$, and $\bar{A}_+ \to - \bar{A}_-$, and $\bar{A}_- \to -\bar{A}_+$.

\subsubsection{Supplementary case}

In the supplementary case there are always a finite number of zeros of $\chi$, and again we do not invert exactly.  Assuming $\hp>1/2$, the resonances are at 
\begin{equation}\label{eq:spec supp}
    \omega_n=-\frac12 \exp\left(\frac{\ln\abs{\rat}}{1-2\hp}\right)\Bigg[
     \sin\left(\frac{2\pi n-\arg(\rat)}{2\hp-1}\right)-i\cos\left(\frac{2\pi n-\arg(\rat)}{2\hp-1}\right),
   \Bigg]
\end{equation}
where
\begin{equation}
n_{-} \leq n \leq n_{+}, \quad \text{where}\quad \displaystyle n_{\pm}=\text{floor}
\left(\frac{\arg(\rat)\pm\pi(2\hp-1)}{2\pi}\right)
%n_{\rm min} \leq n \leq n_{\max}, \quad \text{where}\quad \displaystyle n_{\rm max/min}=\text{floor}\left( \frac{\arg(\rat)\pm\pi(2\hp-1)}{2\pi} \right).
%    \text{floor}\left( \frac{\arg(\rat)-\pi(2\hp-1)}{2\pi} \right)\leq  n \leq \text{floor}\left( \frac{\arg(\rat)+\pi(2\hp-1)}{2\pi} \right).
\end{equation}
is imposed by our restriction to $-\pi/2 <\arg \omega < 3 \pi/2$.
Evidently, a mode instability arises when
\begin{equation}
    2\pi n-\frac{\pi}{2}\left(2\hp-1\right)<\arg\left(\rat\right)<2\pi n+\frac{\pi}{2}\left(2\hp-1\right), \quad \text{(instability criterion, supplementary case)}.
\end{equation}
We note, however, that supplementary modes in the full geometry will have their late-time behavior dictated by the \textit{Dirichlet} case (see dicsussion in Sec.~\ref{sec:critical tail} below), and will not see this spectrum of modes.

\section{$\SO(2,1)$ near-horizon geometries}\label{sec:SO(2,1) geometries}

A near-horizon limit can be defined for any degenerate Killing horizon, giving rise to a universal form for all near-horizon geometries \cite{Reall:2002bh,Astefanesei:2006dd,Kunduri:2007vf}.  When field equations are imposed, the form simplifies further \cite{Kunduri:2013ana}.  In particular, we can encompass all known Einstein-Maxwell solutions, as well as all known restrictions on solutions, with the general ansatz
\begin{subequations}\label{eq:ansatz}
\begin{align}
    ds^2 & = 
    L^2(y) d\hat s^2 
    + \gamma_{IJ}(y)\left(d\phi^I+k^I \hat A \right) \left(d\phi^J + k^J\hat A \right) 
    + \sigma_{ij}(y) dy^i dy^j, \label{eq:KK metric} \\
        A & = Q_I(y) \left( d\phi^I + k^I \hat A \right), \label{eq:KK A}
\end{align}
\end{subequations}
where $d\hat{s}^2$ and $\hat{A}$ are given in Eq.~\eqref{eq:ads2 stuff}.\footnote{Our conventions for $\hat A$ and $k^I$ relate to those in Durkee and Reall \cite{Durkee:2010ea} by $\hat A=-\hat A_{\mathrm{DR}}$ and $k^I=-k^I_{\mathrm{DR}}$.}  
The $\phi^I$ form azimuthal angles ($\phi^I \sim \phi^I+2\pi$) in $N$ orthogonal planes.  We use capital roman letters $I,J,K,\dots$ for these coordinates.  The remaining $d-N-2$ coordinates are denoted $y^i$, with indices $i,j,k,\dots$.  The $k^I$ are constants, while $L$, $\gamma_{IJ}$, $\sigma_{ij}$, and $Q_I$ are functions of $y^i$ that are determined by the field equations.  This ansatz trivially generalizes those of \cite{Figueras:2008qh,Chow:2008dp,Durkee:2010ea,Kunduri:2013ana}.

We will use the language of fiber bundles.  The total space is the $d$-dimensional manifold, with base space $(t,x)$ and fiber $(y^i,\phi_I)$.  We regard $\hat{g}$ and $\hat{A}$ as the metric and gauge field on the base space.  For the fiber, we assign a metric $\mathring{g}$ and gauge field $\mathring{A}$ given by pullback to constant $x$ and $t$ surfaces, 
\begin{align}\label{eq:fiber}
    d\mathring{s}^2 = \gamma_{IJ} d\phi^I d\phi^J + \sigma_{ij} dy^i dy^j, \qquad \mathring{A} = Q_I d\phi^I.
\end{align}
When necessary, we will use Greek mid-alphabet indices $\mu,\nu,\rho,\sigma,\ldots$ to index tensors on the fiber.  We denote the covariant derivative compatible with the fiber metric by $\mathring \nabla_\mu$.

The ansatz \eqref{eq:ansatz} has symmetry group $\SO(2,1)\times U(1)^N$, with the factors associated with the base space and fiber, respectively.  The generators are given by
\begin{align}\label{eq:NHGkilling}
 W_I = \pd_{\phi^I}, \qquad H_0 = \hat{H}_0, \qquad H_+ &= \hat{H}_+ , \qquad H_-  = \hat{H}_- - \frac{2}{x} k^I W_I. %- \frac{2}{x} k^I W_I 
\end{align}
These satisfy the commutation relations $ [ H_+,H_-] = 2 H_0$ and $[H_{\pm},H_0]=H_{\pm}$ of $\SO(2,1)$, with each $U(1)$ generator $W_I$ commuting with everything. 

The coordinates $(t,x,y^i,\phi^I)$ do not extend to the horizon $x=0$. The metric and gauge field are regular, however, in ``ingoing'' coordinates
\begin{equation}\label{eq:ingoing}
    v = t-1/x, \quad \varphi^I = \phi^I - k^I \ln x.
\end{equation}
This transformation leaves the fiber \eqref{eq:fiber} invariant while inducing the coordinate and gauge change \eqref{eq:ads2-ingoing} on the base space \eqref{eq:ads2 stuff}.  That is, after making the transformation \eqref{eq:ingoing}, the metric takes the same form \eqref{eq:ansatz} with $\phi^I \rightarrow \varphi^I$ and $\hat{A} \rightarrow \hat{A}'$ and the hatted quantities now expressed in ingoing coordinates \eqref{eq:ads2-ingoing}.  The Killing fields \eqref{eq:NHGkilling} transform as
\begin{align}\label{eq:NHGkillingin}
 W_I = \pd_{\varphi^I}, \qquad H_0 = \hat{H}_0+k^I W_I , \qquad H_+ &= \hat{H}_+, \qquad H_-=\hat{H}_-+2v\,k^IW_I. 
\end{align}
where Eq.~\eqref{eq:ingoing H H H powah} gives the $\adst$ Killing fields in ingoing coordinates.  Notice how the relationships between hatted and unhatted Killing fields in Eqs.~\eqref{eq:NHGkilling} and \eqref{eq:NHGkillingin} correspond to the generalized Killing field pairings in Eqs.~\eqref{eq:symm bnd} and \eqref{eq:symm hor}, respectively.

\subsection{Charged scalar fields}\label{sec:charged scalar}
Consider now a charged massive (complex) scalar field $\Phi$ in the near-horizon geometry satisfying 
\begin{equation}\label{eq:boxPhi}
  \left(D^2 - \mu^2 \right) \Phi = 0, \qquad D_a = \nabla_a - i e A_a.
\end{equation}
The wave equation separates under the mode ansatz \cite{Durkee:2010ea} 
\begin{equation}\label{eq:Phi ansatz}
    \Phi = \hat\psi(t,x) Y(\phi^I, y^i), \qquad Y=e^{im_I\phi^I}P(y^i). 
\end{equation}
The $Y$ functions satisfy a self-adjoint elliptic equation on the fiber,
\begin{equation}\label{eq:Y eq}
   \mathring{D}^\mu \left(L^2 \mathring{D}_\mu Y\right) + \left( \mathcal{E} - L^2 \mu^2 \right) Y =0,
\end{equation}
where $\mathring{D} = \mathring \nabla - i e \mathring{A}$ is the fiber covariant derivative.  For a compact fiber, the operator appearing in \eqref{eq:Y eq} is self-adjoint with respect to the natural $L^2$ inner product on the fiber.  This guarantees a complete, orthogonal set of eigenfunctions labeled by a discrete set of eigenvalues $m^I$ and $\mathcal E$.  For a non-compact fiber there will generally be a continuous spectrum of allowed values for $\mathcal E$.  In what follows we will assume the fiber is compact, but the analysis of individual modes applies more generally, and results can typically be converted to non-compact cases by  exchanging sums for integrals in the usual way.  An example is given in App.~\ref{sec:planar RN-AdS}. 

The equation for the field $\hat\psi(t,x)$ is
\begin{align}\label{eq:psi ads2 eq}
     \left( \hat D^2 - \hat{\mu}^2 \right) \hat\psi = 0 \qquad \textrm{ with } \qquad \hat{\mu}^2 = \mathcal{E}, \quad \hat{e} = k_I m^I,
\end{align}
where $\hat{D} = \hat \nabla - i \hat{e} \hat{A}$ is the gauge covariant derivative on the $\adst$ base space.  That is, $\psi(t,x)$ obeys the equation for a charged scalar on $\adst$, with the eigenvalues $m^I$ and $\mathcal{E}$ setting the effective mass and charge.\footnote{For the benefit of the reader, we note that our definition for the $\adst$ effective scalar field mass in this section differs from that originally used by Durkee and Reall \cite{Durkee:2010ea}: $\hat\mu^2=\hat\mu^2_{\rm DR}-\hat{e}^2$.}  The solutions of this equation were studied in Sec.~\ref{sec:base space}.  From Eq.~\eqref{eq:hpm}, the exponents $\hpm$ are given by
\begin{align}\label{eq:hpm NHG}
    \hpm = 1/2\pm\sqrt{1/4+\mathcal{E}-(k_I m^I)^2}.
\end{align}
Expressing a mode \eqref{eq:Phi ansatz} in regular coordinates \eqref{eq:ingoing} gives
\begin{align}\label{eq:newPhi}
    \Phi = \hat{\psi}'(v,x) Y(\varphi^I,y^i), \qquad \hat{\psi}' = e^{i m_I k^I \ln x} \hat{\psi},
\end{align}
showing how the coordinate change \eqref{eq:ingoing} properly induces the gauge change \eqref{eq:ads2-ingoing} on the $\adst$ field $\hat{\psi}$ with charge $\hat{e}=k_I m^I$.

\subsection{Gravitational and electromagnetic perturbations}\label{sec:Hertz pots}

 Gravitational and electromagnetic perturbations of vacuum black holes also satisfy decoupled equations in the near-horizon geometry \eqref{eq:ansatz} \cite{Durkee:2010ea}.  One introduces a ``tensor Hertz potential'' $\mathbf\Upsilon_b=\Upsilon_{\mu_1 \dots \mu_{|b|}}$ with $|b|$ indices living on the fiber, where $b=-1,-2$ for electromagnetic and gravitational perturbations, respectively.  The Hertz potential is constructed from a null basis for spacetime, whose real null directions $\ell$ and $n$ are chosen to be 
\begin{subequations}\label{eq:ell and n}
\begin{align}
   n  &= \frac{1}{L\sqrt{2}} \left(x\pd_x-\frac{1}{x}\pd_t + k^I\pd_{\phi^I}\right),\qquad \ell = \frac{1}{L\sqrt{2}} \left(x\pd_x+\frac{1}{x}\pd_t-k^I\pd_{\phi^I}\right).
\end{align}
\end{subequations}
The field equation for the Hertz potential separates under the ansatz 
\begin{equation}\label{eq:GHP scalar}
    \Upsilon_{\mu_1\ldots\mu_{\abs{b}}} = \hat{\psi}_b(t,x)  Y_{\mu_1\ldots\mu_{\abs{b}}}(\phi^I, y^i), \qquad Y_{\mu_1\ldots\mu_{\abs{b}}}=e^{im_I \phi^I} P_{\mu_1\ldots\mu_{\abs{b}}}(y^i).
\end{equation}
The angular eigentensors $Y_{\mu_1 \dots \mu_{\abs{b}}}$ satisfy 
self-adjoint elliptic equations (Eqs. (2.20) and (2.29) of \cite{Durkee:2010ea}; see also Eqs.~(72) and (100) of \cite{Hollands:2014lra}) and are suitably orthogonal and complete \cite{Durkee:2010ea}, with real eigenvalues $\mathcal{E}_b$.  As in the case of scalar perturbations, the $\hat{\psi}_b$ satisfy the $\adst$ charged, massive scalar wave equation \cite{Durkee:2010ea},\footnote{Our convention for the effective electric charge $\hat{e}$ relates to that of Durkee and Reall \cite{Durkee:2010ea} by complex conjugation: $\hat e = \hat e_{\rm DR}^*$. This results from our choice of fields with boost weight $b<0$ for the Hertz potentials.}
\begin{equation}\label{eq:psib eq}
    \left(\hat D^2 - \hat \mu^2 \right)\hat{\psi}_b = 0  \qquad \textrm{ with } \qquad \hat \mu^2 = \mathcal E_b + \hat e^2,
    \quad \hat e =k_I m^I - i b,
\end{equation}
which may be compared with Eq.~\eqref{eq:psi ads2 eq} for the charged scalar.  
From Eq.~\eqref{eq:hpm}, the exponents $\hpm$ are given by
\begin{align}\label{eq:hmp hertz}
    \hpm &=1/2 \pm \sqrt{1/4+\hat\mu^2-\hat e^2}, \no \\ &= 1/2 \pm \sqrt{1/4+\mathcal E_b}.
\end{align}
Notice that while the effective mass and charge can each be complex, the combination appearing under the square root is real.  This ensures that the weight $h$ has the same general properties of the scalar case [Tab.~\ref{tab:h}].  

The null basis \eqref{eq:ell and n} is not regular on the future horizon, since the vector $n$ vanishes there, while the vector $\ell$ blows up.  This can be fixed by rescaling the vectors as
\begin{align}\label{eq:ell and n ingoing}
    \ell \to \ell'= x \ell, \qquad n \to n' = x^{-1} n.
\end{align}
These rescaled vectors are given in ingoing coordinates \eqref{eq:ingoing} by
\begin{align}
       n'&= \frac{1}{L\sqrt{2}} \pd_x,\qquad \ell' = \frac{1}{L\sqrt{2}} \left(x^2\pd_x+2\pd_v-2 x k^I\pd_{\varphi^I}\right),
\end{align}
revealing that, on the horizon, $\ell'$ is tangent to the generators, while $n'$ is transverse.  The precise form of \eqref{eq:ell and n ingoing} is chosen so that the simultaneous change of coordinates \eqref{eq:ingoing} and null basis \eqref{eq:ell and n ingoing} corresponds in $\adst$ to the change \eqref{eq:ads2-ingoing} to horizon-adapted coordinates and $U(1)$ gauge.  In particular, the Hertz potential has boost-weight $b$ [meaning that $\mathbf\Upsilon_b \to \mathbf\Upsilon_b' = x^b \mathbf\Upsilon_b$ under \eqref{eq:ell and n ingoing}], so a mode \eqref{eq:GHP scalar} of $\mathbf\Upsilon_b$ becomes 
\begin{align}\label{eq:newUpsilon}
    \mathbf\Upsilon_b \to \mathbf\Upsilon_b'= \hat{\psi}'_b(v,x) Y(\varphi^I,y^i), \qquad \hat{\psi}'_b = e^{i\left(m_I k^I - ib \right)\ln x }\hat{\psi}_b,
\end{align}
where we now suppress fiber indices on $Y$.  That is, the $\adst$ field $\hat{\psi}_b$ properly transforms as a complex scalar field of charge $\hat{e}=k_I m^I-ib$ under the simultaneous change of coordinates \eqref{eq:ingoing} and null basis \eqref{eq:ell and n ingoing} for the Hertz potential.  Eq.~\eqref{eq:newUpsilon} may be compared with Eq.~\eqref{eq:newPhi}, to which it reduces when $b=0$.

Yet another change of null basis is useful for understanding the decay of electromagnetic and gravitational perturbations.  The original basis \eqref{eq:ell and n} possessed the convenient property that each of its members is invariant under dilations by $H_0$.  This symmetry was destroyed by the change \eqref{eq:ell and n ingoing}, but can be restored by a further time-dependent rescaling as
\begin{align}\label{eq:ell and n ingoing time-dependent}
    \ell''= v \ell' = xv \ell, \qquad n'' = v^{-1} n' = (xv)^{-1} n.
\end{align}
The new legs $\ell''$ and $n''$ are dilation invariant (Lie-derived by $H_0$) as well as regular on the future horizon.  The new $\mathbf\Upsilon_b''$ is given by
\begin{align}\label{eq:double your primes}
    \mathbf\Upsilon_b''= v^b\mathbf\Upsilon_b'.
\end{align}
Note that $\mathbf\Upsilon_b''$ does \textit{not} correspond to some $\hat{\psi}_b''$ on $\adst$ (i.e. there is no analog of \eqref{eq:newUpsilon}), since we rescale the null basis without the corresponding change of coordinates needed to ensure the proper $U(1)$ gauge transformation of $\hat{\psi}_b$.  However, we may still discuss the symmetry properties of $\mathbf\Upsilon_b''$ as a field in the $SO(2,1)$ near-horizon geometry, and these will play a key role in our discussion of the electromagnetic and gravitational perturbations constructed from  $\mathbf\Upsilon_b''$ in Sec.~\ref{sec:EMGrav} below.

\section{Full Geometries}\label{sec:full space}

Now consider a spacetime and electromagnetic field geometry whose near-horizon limit takes the form \eqref{eq:ansatz}.  That is, we consider a stationary, axisymmetric\footnote{That is, we suppose there are $N$ commuting Killing fields with closed orbits.  Every stationary black hole has at least one such Killing field \cite{Hollands:2006rj}.} Einstein-Maxwell solution, denoted $\tilde{g},\tilde{A}$, and suppose that there exist coordinates $( t, x,y^i,\phi^I)$ together with a $U(1)$ gauge choice such that limit 
\begin{align}\label{eq:limitaceous}
    \qquad   x \to 0 \textrm{ \ fixing \ }  x  t \qquad \textrm{(near-horizon limit)}
\end{align}
recovers the geometry \eqref{eq:ansatz}.\footnote{More formally, the limit is given by changing coordinates to $T=\lambda t$ and $X =x/\lambda$ and letting $\lambda \to 0$.}  This fixes the free functions and parameters in the ansatz; see Sec.~\ref{sec:examples} below for an example.  We first present the case of scalar perturbations and then discuss electromagnetic and gravitational perturbations in Sec.~\ref{sec:EMGrav}.  A charged, massive scalar field in the full geometry satisfies
\begin{align}\label{eq:exterior wave eqn}
    \left( \tilde{D}^2 - \mu^2 \right) \Phi = 0, \qquad \tilde{D} = \tilde{\nabla} - i e \tilde{A},
\end{align}
where $\tilde{\nabla}$ is the metric-compatible derivative on the full geometry, making $\tilde{D}$ the gauge-covariant derivative.  We will require that mode solutions of this equation suitably match onto those of the near-horizon geometry.  

\subsection{Separable case}

We will first consider the case where the equation separates on the full geometry, by which we mean that the mode ansatz
\begin{equation}\label{eq:Phi ansatz full}
    \Phi = e^{- i \omega t} \tilde{R}(x) \tilde{Y}(\phi^I, y^i),
\end{equation}
gives rise to an ordinary differential equation for $\tilde{R}(x)$ and an elliptic (self-adjoint) PDE for the angular functions $\tilde Y$.  These equations will in general depend on $\omega$ and $m_I$ as well as an additional separation constant $\tilde{\mathcal{E}}$.

The radial functions $\tilde{R}$ are assumed to satisfy a linear second-order ODE.  In a given problem, boundary conditions will be imposed at $x=0$ (the horizon) as well as at larger $x$ (typically at asymptotic infinity).  We denote the corresponding solutions by ``in'' and ``$\infty$''
\begin{subequations}\label{eq:ininf}
\begin{align}
    \tilde{R}_{\rm in}(x)&: \textrm{ satisfies ingoing boundary conditions as $x \to 0$}\label{eq:Rin def}, \\
   \Rfinf(x) &: \textrm{ satisfies boundary conditions at some $x>0$ (typically $x \to \infty$)}. \label{eq:Rup def}
\end{align}
\end{subequations}
Below we will make the notion of ingoing boundary conditions precise by demanding a match to the $\adst$ ingoing solution [Eq.~\eqref{eq:Rnear limits to Ads2}].  On the other hand, we  leave $\Rfinf(x)$ arbitrary to cover a general choice of boundary conditions.

To consider solutions arising from initial data, we introduce a Green function $\tilde G$, which satisfies 
\begin{align}\label{eq:tildeG wave eq}
\left( \tilde{D}^2 - \mu^2 \right) \tilde G = \delta_d,
\end{align}
where $\delta_d$ is the invariant delta-function in $d$-dimensions.  We  mode decompose $\tilde G$ as
\begin{align}\label{eq:tilde G}
   \tilde G=\frac{1}{2\pi}\int_{-\infty+ic}^{\infty+ic} \sum_{L} e^{-i\omega(t-t')}\tilde Y_L(\phi^I, y^i)\tilde Y_{L}^*(\phi'{}^I,y'{}^i)\tilde g_{L}(x,x')\,d\omega.
\end{align}
Here we introduce the notation $L$ for the collection $\{\tilde{\mathcal E}, m^I\}$ indexing the eigenvalues.  Each $\tilde{g}_{L}$ serves as the transfer function for its mode. In what follows we will usually suppress the index $L$.  The transfer functions satisfy equations of the form
\begin{align}\label{eq:tildeg eqn}
    \left(\pd_x^2+\beta(x,\omega)\pd_x+\gamma(x,\omega)\right)\tilde g(x,x')=f(x')\delta(x-x').
\end{align}
Importantly, the function $f(x')$ is independent of $\omega$ since it follows from the principal (highest-derivative) part of $\tilde{D}^2$ together with metric factors in the coordinate expression of $\delta_d$.  The homogeneous solutions are just the $\tilde{R}$ as already defined.  To properly give causal dynamics, the transfer function must agree with the in solution for $x<x'$ and with the $\infty$ solution for $x>x'$.  Matching at the support of the delta function gives 
\begin{align}\label{eq:tilde g}
 \tilde g(x,x')=f(x')\frac{\Rfin(x_<)\Rfinf(x_>)}{W[\Rfin,\Rfinf]},
\end{align}
where $W[\Rfin,\Rfinf]$ denotes the Wronskian of $\Rfin$ and $\Rfinf$, evaluated at $x'$.

\subsection{Critical frequency}\label{sec:crit freq}
The near-horizon limit was given above as $x \to 0$ fixing $xt$.  Working in frequency space, the relevant notion of near-horizon limit is $x \to 0$ fixing $x/\omega$.  This choice preserves the $e^{-i \omega t}$ factor in the mode ansatz \eqref{eq:Phi ansatz full} as $x \to 0$ fixing $tx$, ensuring that the limiting eigenfunctions satisfy the appropriate equations in the near-horizon geometry.  We may thus equivalently phrase the near-horizon limit as $\omega \to 0$ fixing $x/\omega$, showing an association between the near-horizon geometry and the critical frequency $\omega=0$.\footnote{The critical frequency is always zero in the coordinates we have chosen.  In specific metrics, it may be non-zero in the common coordinate systems.  For example, in Boyer-Lindquist coordinates for Kerr, $\omega=m\Omega_H$ where $m$ is the azimuthal number of the mode and $\Omega_H$ is the horizon frequency.  This is more commonly known as the superradiant bound frequency.  See Sec.~\ref{sec:examples} for a detailed analysis of extremal Kerr-Newman.}  To analyze the full behavior near the critical frequency, we will also need a second, ``far'' limit of $\omega \to 0$ fixing $x$.  That is, we use the terminology:
\begin{subequations}
\begin{align}\label{eq:limits}
    \textrm{near limit: \ } \omega & \to 0 \textrm{ \ fixing \ } x/\omega \\
    \textrm{far limit: \ } \omega &\to 0 \textrm{ \ fixing \ } x.
\end{align}
\end{subequations}
The distinction between near and far is irrelevant for the angular eigenfunctions $\tilde{Y}$, which do not depend on $x$.  Since these will limit to solutions of the near-horizon elliptic equation \eqref{eq:Y eq}, we can make them match for each index by requiring 
\begin{align}\label{eq:Y limit}
    \lim_{\omega \to 0} \tilde{Y}(\phi^I,y^i) = Y(\phi^I,y^i).
\end{align}
For the radial functions we must consider both near and far limits, using the method of matched asymptotic expansions.  We will assume the following properties:\footnote{\label{2h ass} In the special case where $2\hp$ is an integer, the overlap region behavior $x^{-\hp}$ should be replaced with $\ln(x)x^{-h_+}$ in the assumptions below.}
\begin{itemize}
    \item (i) The far limit of the radial equation has linearly independent solutions $\tilde{R}^{\rm far}_{+}$ and $\tilde{R}^{\rm far}_{-}$ 
    satisfying
\begin{subequations}
   \begin{align}
        \tilde{R}^{\rm far}_{+}&\sim x^{-\hp}, \qquad x \to 0, \\
        \tilde{R}^{\rm far}_{-}&\sim x^{-\hm}, \qquad x \to 0.
    \end{align}
\end{subequations}
    \item (ii) The far limit of the $\infty$ function solves the far limit of the radial equation, 
    \begin{align}\label{eq:Rfarup}
    \Rtfinf:=\lim_{\substack{\omega \to 0 \\ \textrm{fix } x}}\tilde{R}_{\infty} = B_+ \tilde{R}^{\rm far}_{+} + B_- \tilde{R}^{\rm far}_{\rm -}.
    \end{align} 
    \vspace{-.5cm}
    \item (iii) The near limit of the in function is equal to the $\adst$ in function,
    \begin{align}\label{eq:Rnear limits to Ads2}
    \tilde{R}^{\rm near}_{\rm in} := \lim_{\substack{\omega \to 0 \\ \textrm{fix } x/\omega}} \tilde{R}_{\rm in}  = \hat R_{\rm in}.
    \end{align}
    \vspace{-.65cm}
    \item (iv) Everything matches appropriately,
    \begin{subequations}
     \begin{align}\label{eq:matching conds}
    \tilde{R}_{\rm in}^{\rm near} & \sim A_+ x^{-\hp} + A_- x^{-\hm}, \qquad x \to \infty, \\
    \tilde{R}_{\rm in}^{\rm far} &  \sim A_+ x^{-\hp} + A_- x^{-\hm}, \qquad x \to 0, \label{eq:in far matching} \\
    \tilde{R}_{\infty}^{\rm near} &  \sim B_+ x^{-\hp} + B_- x^{-\hm}, \qquad x \to \infty, \\
    \tilde{R}_{\infty}^{\rm far} &  \sim B_+ x^{-\hp} + B_- x^{-\hm}, \qquad x \to 0. \label{eq:up far matching}
\end{align}
    \end{subequations}
\end{itemize}
In these assumptions, $\hp, \hat{e}$, and $\hat{\mu}$ are to be computed from the near-horizon eigenvalues $\mathcal{E}$ and $m^I$ using Eqs.~\eqref{eq:psi ads2 eq} and \eqref{eq:hpm NHG}.  The effective $\adst$ boundary conditions $\NO:=B_+/B_-$ \eqref{eq:N} can be determined from the small-$x$ behavior of the far $\infty$ function, Eq.~\eqref{eq:up far matching}.

The above assumptions are not all independent, including significant redundancy in order to establish notation.  Assumption (i) guarantees that the far limit can match on to the near limit by demanding the appropriate asymptotic behaviors of $x^{-\hp}$ and $x^{-\hm}$.  Assumption (ii) ensures that the $\infty$ function properly limits to a solution of the limiting radial equation.  Assumption (iii) ensures that the in function properly limits to the in solution in the near-horizon geometry.  Finally, assumption (iv) ensures that each of the in and $\infty$ solutions properly matches according to the method of matched asymptotic expansions. In particular, we can write the far in function as
\begin{align}\label{eq:Rfarin}
    \tilde{R}^{\rm far}_{\rm in} = A_+ \tilde{R}^{\rm far}_{+} + A_- \tilde{R}^{\rm far}_{-},
\end{align}
where $A_+$ and $A_-$ are given in Eq.~\eqref{eq:Apm}.  

\subsection{Critical tail}\label{sec:critical tail}

By following a standard sequence of steps (done, e.g., for the Kerr spacetime in \cite{Gralla:2017lto}), the solution arising from initial data supported away from the horizon can be reduced to a sum over convolutions of initial data modes with the transfer function for each mode.  Thus transfer functions capture the generic behavior of the field in the sense that features in the transfer functions will generically manifest in the field in a calculable way.  We will study the behavior of the transfer functions near the critical frequency $\omega=0$.  If the full transfer function has no other singular points in the complex plane at equal or larger (real part of) frequency, the results will correspond to the late-time behavior of the field.  More generally, we can still expect the behavior near $\omega=0$ to show up in some way during the evolution, much as simple poles (quasinormal modes) are typically seen as identifiable damped oscillations at intermediate times.  Here we will focus on the $\omega \to 0$ limit of the transfer functions, whose associated behavior we call the ``critical tail'' in light of the association with critical phenomena \cite{Faulkner:2009wj,Ren:2012hg,Gralla:2017lto}.  More precisely, we define the critical tail $\tilde{G}_{\rm tail}(t)$ of each mode to be the inverse Laplace transform of the leading non-analytic term in the $\omega \to 0$ expansion of the transfer function $\tilde g$.\footnote{If there are infinitely many poles as $\omega \to 0$ [case II of the classification \eqref{eq:principal cases}], we avoid the term ``tail'' as the behavior will not be a simple power law.}  There are different tails for the far and near regions, which we treat separately below.  

As in the case of pure $\adst$, for computation purposes we assume that $2\hp$ is not an integer; however, the results remain valid in that limit, so we effectively treat $2\hp \in \mathbb{Z}$ by analytic continuation (see footnotes \ref{Whitlog} and \ref{2h ass}).  We further assume that the mode is not discrete [see Tab.~\ref{tab:h}]; this case is discussed in App.~\ref{sec:pain in the ax}.

\subsubsection{Far region and off-horizon tail}\label{sec:crit tail off}

We begin by considering the field off of the horizon ($x>0$), whose small-$\omega$  features correspond to the ``far'' limit $\omega \to 0$ fixing $x$.  The initial data is assumed to be supported away from the horizon, and hence we similarly consider $x'$ to be in the far region.  The relevant limit of the transfer function \eqref{eq:tilde g} is thus
\begin{align}\label{eq:gff def}
 \gff(x,x') = f(x') \frac{\Rtfin(x_<) \Rtfinf(x_>)}{W[\Rtfin,\Rtfinf]},
\end{align}
where ff stands for ``far far'' and where the Wronskian $W$ is computed at the point $x'$.  The $\omega$-independent function $f(x')$ was introduced in Eq.~\eqref{eq:tildeg eqn}. 
Using Eqs.~\eqref{eq:Rfarin}, \eqref{eq:Rfarup}, \eqref{eq:barred quantities}, we may express $\gff$ as
\begin{align}\label{eq:dirty pony}
    \gff(x,x')=\frac{f(x')}{W[\Rtfp,\Rtfm]} \frac{\left( (-2i\omega)^{2\hp-1}\GO \tilde{R}^{\rm far}_{+}(x_<)+\tilde{R}^{\rm far}_{-}(x_<) \right) \left( 
    \NO \tilde{R}^{\rm far}_{\rm +}(x_>)+\tilde{R}^{\rm far}_{\rm -}(x_>) \right)}{\NO-\GO (-2i\omega)^{2\hp-1}}.
\end{align}
In Eq.~\eqref{eq:dirty pony}, the $\omega$-dependence is completely explicit: $f$, $\GO$, $\NO$ and $\tilde{R}^{\rm far}_{\pm/\infty}$ are all independent of $\omega$.  Alternatively, we can repackage the $\omega$-dependence in terms of the Dirichlet boundary-boundary correlator \eqref{eq:g pd pd script G} of near-horizon holography,
\begin{align}\label{eq:holographic pony}
    \gff(x,x')=\frac{f(x')}{W[\Rtfp,\Rtfm]} \frac{\left( (2h_+-1)\hat{g}^{+}_{\pd \pd}(\omega) \tilde{R}^{\rm far}_{+}(x_<)+\tilde{R}^{\rm far}_{-}(x_<) \right) \left( 
    \NO \tilde{R}^{\rm far}_{\rm +}(x_>)+\tilde{R}^{\rm far}_{\rm -}(x_>) \right)}{\NO-(2h_+-1)\hat{g}^{+}_{\pd \pd}(\omega)}.
\end{align}

\paragraph{{Supplementary modes}}\label{sec:ff supp}

In the case of supplementary modes (see Tab.~\ref{tab:h}), $\hp>1/2$ so that $\hat{g}_{\pd \pd}^+(\omega)\sim \omega^{2\hp-1}$ is small as $\omega \to 0$. This makes the far-far transfer function \eqref{eq:holographic pony} take the form
\begin{align}\label{eq:ff small omega}
    \tilde{g}^{\rm ff} \sim a(x,x') + b(x,x') \hat{g}_{\pd \pd}^+(\omega) , \qquad \omega \to 0,
\end{align}
where the 
functions $a$ and $b$ are independent of $\omega$.  The first term does not contribute at late times (it is a ``contact term''). The second term is the leading non-analytic contribution for small $\omega$, which defines the critical tail.  The inverse Laplace transform of this term is just the boundary-boundary retarded correlator \eqref{eq:Gpdpd}, giving the tail as 
\begin{align}\label{eq:Gff large t sup}
  \Gff_{\rm tail} (t,x,x') & = b(x,x') \hat G^+_{\pd \pd}(t) \propto t^{-2\hp}. 
\end{align}
Here and below, $\propto$ means equality up to multiplication by a function of the coordinates not displayed (in this case $x$ and $x'$).  The factor $b(x,x')$, which may be straightforwardly determined from Eq.~\eqref{eq:holographic pony},
encodes the non-universal physics of the far region.  The time-dependence is universal.

\paragraph{Principal modes}\label{sec:ff prince}

For principal modes where $\hp=1/2+ir$ with $r>0$ (see Tab.~\ref{tab:h}), $\hat{g}_{\pd \pd}^+(\omega)\sim \omega^{2\hp-1}=\omega^{2ir}$ is a logarithmic phase and the transfer function \eqref{eq:dirty pony} does not simplify as $\omega \to 0$.  However, it has the same analytic structure as the boundary-boundary mixed transfer function given previously in Eq.~\eqref{eq:hat sf g x xp large}, so we may take advantage of previously derived results in $\adst$ (Sec.~\ref{sec:mixed special BCs}) to obtain the inverse Laplace transform.  The results depend upon the ratio $\GO/\NO$ and the imaginary part of $\hp$, denoted $r$.  The unstable range of Eq.~\eqref{eq:QNM instability crit prince} corresponds in the present small-$\omega$ context to an infinite number of poles along a line as $\omega \to 0$, which has been seen in previous work \cite{Faulkner:2009wj, Hartnoll:2016apf} to correspond to a condensate-type (``superconducting'') instability of the spacetime,
\begin{equation}\label{eq:instability condense}
 \textrm{Range for condensate instability: \ \ } e^{-\pi r}<\abs{\rat}<e^{\pi r}. 
\end{equation}
Outside of this range, the dynamics associated with $\omega \to 0$ will be stable.  In the range $e^{\pi r} < |\GO/\NO| <e^{3\pi r}$ there are stable poles with spectrum given in Eq.~\eqref{eq:spec prince}.  It is presently unclear what the associated dynamics are in this case.  In the remaining ranges $|\GO/\NO| < e^{-\pi r}$ and $|\GO/\NO| > e^{3 \pi r}$  (cases I and III, respectively), the inverse Laplace transform of \eqref{eq:dirty pony} may be determined from Eq.~\eqref{eq:G xx' big 1} as
\begin{align} \label{eq:Gff large t princ}
  \Gff_{\rm tail}(t,x,x')& = t^{-1} \sum_{n=0}^\infty\left( c^\pm_n(x,x') + d^\pm_n(x,x') t^{\pm 2ir}\right) t^{\pm 2i n r},
\end{align}
where the lower and upper signs correspond to cases I and III, respectively. The non-universal coefficients $c^{\pm}_n(x,x')$ and $d^\pm_n(x,x')$, whose precise form is irrelevant for our arguments, may be determined from Eq.~\eqref{eq:G xx' big 1}.  The $1/t$ tail is universal.

\subsubsection{Near region and on-horizon tail}\label{sec:horizon tail}

We now consider the field on the horizon ($x=0$), whose small-$\omega$  features are visible in the ``near'' limit $\omega \to 0$ fixing $x/\omega$.  Again, the initial data is assumed to be supported away from the horizon, and we thus consider $x'$ to be in the far region.  The relevant limit of the transfer function  \eqref{eq:tilde g} in this case is given by
\begin{align}\label{eq:gnf def}
 \gnf(x,x') = f(x') \frac{\Rtnin(x) \Rtfinf(x')}{W[\Rtfin,\Rtfinf]},
\end{align}
where nf stands for ``near far''. As usual, $W$ is computed at the point $x'$, explaining the appearance of the far functions.  Crucially, this means that $W$ is independent of frequency.   Using Eqs.~\eqref{eq:Rin from Rp Rm} and \eqref{eq:Rfarup}, we may write 
\begin{align}\label{eq:gnf2}
 \gnf(x,x') = \frac{f(x')}{W[\Rtfp,\Rtfm]}\frac{\left((-2i\omega)^{2\hp-1}\GO\hRp(x_<)+\hRm(x_<)\right)
  \left(\NO \tilde{R}^{\rm far}_{\rm +}(x_>)+\tilde{R}^{\rm far}_{\rm -}(x_>)\right)}{(1-2\hp)\left(\mathcal N_0-\GO(-2i\omega)^{2\hp-1}\right)}.
\end{align}
This is in fact proportional to the large-$x'$ mixed $\adst$ transfer function \eqref{eq:hat sf g xp large}.
Unlike in the far-far case \eqref{eq:dirty pony}, the $\omega$-dependence in \eqref{eq:gnf2} is not all explicit, since the $\adst$ functions $\hat{R}^\pm$ depend on $\omega$.

\paragraph{Supplementary Modes}\label{sec:crit tail nf supp}

Recall that 
 $\omega^{2\hp-1}$ is a small quantity for supplementary modes.  As $\omega \to 0$ we may the drop the $\omega$-dependent term in the denominator of \eqref{eq:gnf2}, giving
\begin{equation}\label{eq:tilde g nf supp}
    \gnf(x,x') \sim u(x') \hat g^{+}_{\rm B\pd}(x), \qquad \omega \to 0,
\end{equation}
where the precise form of the function $u(x')$ is unnecessary.  That is, the near-far transfer function is proportional to the $\adst$ Dirichlet boundary-bulk propagator as $\omega \to 0$.  In the time domain this is just $\hat{G}_{\rm B \pd}$ as given in \eqref{eq:hat Gpm Bpd} or \eqref{eq:hat Gpm Bpd in time}.  We present the result in ingoing coordaintes,\footnote{Strictly speaking, we have not defined an ingoing coordinate $v$ in the full geometry; here we mean any coordinate $v$ such that $v=t-1/x$ as $x \to 0$ fixing $xt$ (or equivalently $xv$) and $\varphi^I = \phi^I - k^I \ln x$ [see Eq.~\eqref{eq:ingoing}], with modes defined relative to $e^{i m_I \varphi^I}$.}
\begin{align}%p8e19
   \tilde G^{\rm nf}_{\rm tail}(v,x,x') = u(x') \hat{G}^+_{\rm B \pd}(x,v) \propto \left(\frac{v}{2}\right)^{-i\hat e-\hp}\left(1+\frac{vx}{2}\right)^{i\hat e-\hp},\label{eq:oink} 
\end{align}
which may be compared to Eq.~\eqref{eq:Gff large t sup}.  The details of the far region appear in the non-universal function $u(x')$, but the complete near-region behavior (small $x$ and large $v$) is universal.  In particular, the Aretakis instability is present and the on-horizon tail is $v^{-i\hat{e}-\hp}$.  We can emphasize the latter by evaluating on the horizon and expressing in terms of the horizon-boundary correlator \eqref{eq:GHpd},
\begin{align}\label{eq:moo}
    \tilde G^{\rm nf}_{\rm tail}|_{x=0} = u(x') \hat{G}^+_{\rm \mathcal{H} \pd}(v) \propto v^{-i\hat e-\hp}.
\end{align}

\paragraph{Principal modes}\label{}  Recall that $\omega^{2\hp-1}$ is a logarithmic phase for principal modes.  We can no longer drop any terms in Eq.~\eqref{eq:gnf2}, but we can make use of the fact that it is proportional  to the large-$x'$ mixed $\adst$ transfer function \eqref{eq:hat sf g xp large},
\begin{align}
    \tilde g^{\rm nf} = w(x') \hat{g}_{\rm mix}(x,x'\to\infty),
\end{align}
where by $\hat{g}_{\rm mix}(x,x'\to\infty)$ we mean the right-hand side of Eq.~\eqref{eq:hat sf g xp large}.  The inverse Laplace transform was computed in Eq.~\eqref{eq:sf ginvP1}, and we have
\begin{align}\label{eq:hot stuff combo}
    \tilde G^{\rm nf}_{\rm tail}(v,x,x') = \frac{1}{\sqrt{v}} \,\,  y_\pm(x')  \sum_{n=0}^\infty (v/2)^{\pm2irn\pm ir-i\he} z_{\pm n}(vx),
\end{align}
where 
\begin{equation}
    z_n(vx) := (\irat)^n {}_2\tilde{F}_1\left(\hp-i\he,1-\hp-i\he;1-\hp-i\he-(1-2\hp)n;-vx/2\right),
\end{equation}
and where the $-$ sign is for $\abs{\rat}e^{\pi r} < 1$ (case I)
and $+$ sign is chosen for $\abs{\rat}e^{-3\pi r} > 1$ (case III).  Again, the non-universal coefficients $y_\pm$ may be determined from relations given previously.  Although the field \eqref{eq:hot stuff combo} is not precisely self-similar, it is given by a sum of terms, each of which is self-similar with the same real part of scaling exponent (in this case $-1/2$).  This property was called ``weakly self-similar'' in Ref.~\cite{Gralla:2017lto}, where it was shown to still entail the Aretakis instability.

The properties of the hypergeometric function guarantee that Eq.~\eqref{eq:hot stuff combo} interpolates between $1/\sqrt{v}$ decay on the horizon $x=0$ and $1/v$ decay as $x \to \infty$ (reproducing the far-far result of $1/t$).  On the horizon the hypergeometric function is equal to unity, so we have
\begin{align}\label{eq:hot stuff chicken combo}
   G^{\rm nf}_{\rm tail}|_{x=0} = \frac{1}{\sqrt{v}} \,\,  y_\pm(x')  \sum_{n=0}^\infty\left(\frac{\NO}{\GO}\right)^{\pm n} (v/2)^{\pm2irn\pm ir-i\he} ,
\end{align}
where again the $-$ sign is for $\abs{\rat}e^{\pi r} < 1$ (case I)
and $+$ sign is chosen for $\abs{\rat}e^{-3\pi r} > 1$ (case III).
In particular, the on-horizon decay is universally like $1/\sqrt{v}$.

\subsection{Non-separable equations}\label{sec:non seps}
Above we assumed that the wave equation separates in the full geometry, which is rather restrictive.  However, the analysis generalizes straightforwardly to non-separable geometries, as we are mainly interested in the effects of the near-horizon region, where the equations still separate.  Considering a non-separable full geometry merely complicates the analysis of the non-universal factors, as we now explain.

The important point is that the mode decomposition remains well-defined even when the field equations do not separate.  In fact, since we no longer care about separating the equations, we can simplify by choosing $\tilde{Y}=Y$, i.e., we use the near-horizon angular eigenfunctions to decompose the full perturbation.
Adopting the ansatz \eqref{eq:Phi ansatz full} for each angular index $L=\{\mathcal{E},m_1,\ldots,m_N\}$ now results in a coupled set of equations for the radial functions $\tilde R_{L}$, which takes the form 
\begin{equation}\label{eq:R eq ns} 
A^{L'}_{\ L}(x) \pd_x^2 \tilde R_{L'}+B^{L'}_{\ L}(x,\omega)\pd_{x}\tilde 
  R_{L'}+C^{L'}_{\ L}(x,\omega)\tilde R_{L'}=0,
\end{equation}
for some matrix-valued functions $A$, $B$, and $C$, 
with implied summation over repeated indices $L'$. 
We can again define the $\infty$ and in solutions by appropriate boundary conditions, as in Eqs.~\eqref{eq:ininf}. 
Instead of the decomposition \eqref{eq:tilde G} for the Green function, 
we now adopt 
\begin{equation}\label{eq:G ns}
  \tilde G=\frac{1}{2\pi}\int_{-\infty+ic}^{\infty+ic} \sum_{L,L'} e^{-i\omega(t-t')}\tilde Y_L(\phi^I, y^i)\tilde Y_{L'}^*(\phi'{}^I,y'{}^i)\tilde g_{LL'}(x,x')\,d\omega,   
\end{equation}
with transfer functions $\tilde g_{LL'}$ satisfying
\begin{equation}\label{eq:g eq ns}
   A^{L''}_{\ L}(x)\, \pd_{x}^2\, \tilde g_{L''L'}(x,x')+B^{L''}_{\ L}(x,\omega)\,\pd_{x}\tilde g_{L''L'}(x,x')+C^{L''}_{\ L}(x,\omega)\,\tilde g_{L''L'}(x,x')= \epsilon_{LL'}(x')\delta(x-x')
\end{equation}
for some $\omega$-independent functions $\epsilon_{LL'}(x')$ coming from the metric determinant. 
 The solution satisfying the boundary conditions is then
    \begin{align}\label{eq:tilde g non-sep}
    \tilde g_{LL'}(x,x')=f_{LL'}(x')\frac{\tilde{R}^{\rm in}_L(x_<)\tilde{R}^{\rm \infty}_{L}(x_>)}{W[\tilde{R}^{\rm in}_L(x'),\tilde{R}^{\rm \infty}_{L}(x')]},
\end{align}
where $W$ is the Wronskian and  $f_{LL'} = (A^{-1})^{L''}_{\ L}\epsilon_{L''L'}$, assuming $A$ is invertible. As defined, there is no sum over the $L$ indices on the radial functions.   This is the same formula as Eq.~\eqref{eq:tilde g} above, except that now $f(x')$ depends on the indices $L$ and $L'$.    We now make the identical assumptions (i)-(iv) of Sec.~\ref{sec:crit freq} to ensure proper matching of near and far expansions near the critical frequency $\omega \to 0$.  The remaining calculations of the critical tail follow without change. 

\subsection{Electromagnetic and gravitational perturbations}\label{sec:EMGrav}

We treat electromagnetic and gravitational perturbations using the Hertz potential formalism \cite{Cohen:1974,Chrzanowski:1975wv,Wald:1978vm,Stewart1979Recon,Godazgar:2011sn, Hollands:2014lra}.  We choose any stationary, axisymmetric null basis for spacetime that reduces to Eqs.~\eqref{eq:ell and n} in the near-horizon limit.  Although the equations for the Hertz potentials $\mathbf\Upsilon_b$ (fiber indices suppressed as in Sec.~\ref{sec:Hertz pots}) will not separate in the full spacetime, by the same arguments of Sec.~\ref{sec:non seps} we can still use the assumptions and calculations of this section for the purposes of calculating the critical tail.  There is a minor subtlety in translating the results in that different null bases are used on and off the horizon.  In particular, scalar results presented in $(t,x)$ coordinates are promoted to Hertz potential results in the original null basis with near-horizon limit \eqref{eq:ell and n}, while scalar results presented in $(v,x)$ coordinates refer to Hertz potentials in the horizon-regular null basis \eqref{eq:ell and n ingoing}.\footnote{This follows from the fact that the passage to regular coordinates and gauge in $\adst$ corresponds to a simultaneous change of coordinates and null basis in the full geometry; see discussion surrounding Eq.~\eqref{eq:newUpsilon}.}

The qualitatively new feature of the electromagnetic and gravitational cases is the need to construct\footnote{For gravitational perturbations of Kerr, it has been shown that the metric may be constructed from the Hertz potentials up to gauge and non-dynamical degrees of freedom \cite{Wald:1973jmp}.  We expect the same to be true more generally.} the perturbation from the Hertz potential $\mathbf\Upsilon_b$.  Off the horizon, the critical tail of the Hertz potential is a well-behaved function of $(t,x)$ and there is no subtlety: in a suitable gauge, the metric (or electromagnetic) perturbation and all its derivatives will share the same power law decay (or at least decay no slower).  On the horizon, it is convenient to use the near-horizon dilation symmetry to organize the calculation and results \cite{Gralla:2017lto}.  
We first consider the case of a supplementary mode in the primed null basis  \eqref{eq:ell and n ingoing}, for which symmetry considerations (Sec.~\ref{sec:symmetry}, Eqs.~\eqref{eq:GBpd symm} and \eqref{eq:horizon H0}) or direct calculations (Sec.~\ref{sec:critical tail}, Eq.~\eqref{eq:oink}) show that
\begin{align}\label{eq:fun times}
\Lie_{H_0} \mathbf\Upsilon_b' = (- h- b) \mathbf\Upsilon_b',
\end{align}
where $H_0$ is given in ingoing coordinates in Eq.~\eqref{eq:NHGkillingin}.\footnote{Note that in our previous paper \cite{Gralla:2017lto} we defined $H_0=v \pd_v - x \pd_x$ instead of the version $H_0=v \pd_v - x \pd_x + k^I\pd_{\varphi^I}$ used here.}
Here we refer to the critical tail of an individual supplementary mode.  When a tensor $T$ satisfies $\Lie_{H_0} T = p T$ we say that it is dilation self-similar with weight $p$.  Thus the weight of $\mathbf\Upsilon_b'$ is $-h-b$. 

To leverage the dilation symmetry we change to the dilation-invariant (and still horizon-regular) null basis $\eqref{eq:ell and n ingoing time-dependent}$.  This sets $\mathbf\Upsilon_b' \to \mathbf\Upsilon_b''=v^b \mathbf\Upsilon_b'$ [Eq.~\eqref{eq:double your primes}], so that
\begin{align}
\Lie_{H_0} \mathbf\Upsilon_b'' = -h\mathbf\Upsilon_b''. 
\end{align}
Importantly, the dilation-weight $-h$ of $\mathbf\Upsilon_b''$ is independent of the boost weight $b$.  Since $\mathbf\Upsilon_b''$ is defined relative to a dilation-invariant null basis, the perturbation construction procedure preserves the dilation weight \cite{Gralla:2017lto}.  That is, there exists a gauge where the metric perturbation $\delta g$ corresponding to the critical tail of a mode $\mathbf\Upsilon_b''$ satisfies
\begin{align}
    \Lie_{H_0} \delta g = -h \,\delta g,
\end{align}
and similarly for the gauge potential $\delta A$ of a mode of an electromagnetic perturbation.  

The above analysis was for a single supplementary mode.  Although principal modes are not precisely self-similar, they decompose into a sum  [Eq.~\eqref{eq:hot stuff combo}] of self-similar terms of dilation-weight $p_i$, all with the same real part $\textrm{Re}[p_i]=-1/2$.  This property was called \textit{weak} self-similarity in Ref.~\cite{Gralla:2017lto}.  The final metric perturbation (or electromagnetic gauge potential) inherits the weak self-similarity of its constituent modes.  That is, the critical tail of the metric perturbation (in a suitable gauge) is a sum of terms, each of which satisfies
\begin{align}
    \Lie_{H_0} \delta g = p\, \delta g, \qquad \textrm{Re}[p] \leq -1/2.
\end{align}
As explained in more detail in \cite{Gralla:2017lto} (and barring subtleties in the convergence of the sum), the $p$ with the smallest real part sets the decay of any component of any tensor constructed geometrically from the perturbation.  In particular, we may conclude that, despite the growth of transverse derivatives along the event horizon, all scalar invariants decay.

\subsection{Summary}\label{sec:summary}

In this section we have demonstrated the detailed relationship between full-spacetime perturbations near the critical frequency (the ``critical tail'') and corresponding perturbations of $\adst$ with a constant electric field.  We now summarize for the reader's convenience.  For scalars $\Phi$ or Hertz potentials $\mathbf\Upsilon$ in a stationary null basis:\footnote{Results on the horizon refer to a null basis reducing to Eq.~\eqref{eq:ell and n ingoing} in the near-horizon limit, while results off the horizon refer to a static null basis reducing to Eq.~\eqref{eq:ell and n} in the near-horizon limit.}
\begin{itemize}
    \item For supplementary modes off the horizon, the critical tail is given by the Dirichlet boundary-boundary correlator $G^+_{\pd \pd}(t) \propto t^{-2\hp}$ [Eq.~\eqref{eq:Gff large t sup} above].
    \item For supplementary modes on the horizon, the critical tail is given by the Dirichlet horizon-boundary correlator $\hat{G}^+_{\rm \mathcal{H} \pd}(v) \propto v^{-i\hat e-\hp}$ [Eq.~\eqref{eq:moo} above].
    \item For supplementary modes asymptotically near the horizon ($x \to 0$ fixing $xv$), the critical tail is given by the Dirichlet bulk-boundary propagator $\hat{G}^+_{\rm B \pd}(x,v) \propto
   v^{-i\hat e-\hp}\left(1+\frac{vx}{2}\right)^{i\hat e-\hp}$ 
    [Eq.~\eqref{eq:oink} above], whose self-similarity gives rise to the Aretakis instability.
    \item For principal modes, the critical tail is determined by the mixed $\adst$ two-point function.  Provided there are no poles, the critical tail universally decays like $1/t$ [Eq.~\eqref{eq:Gff large t princ}] and $1/\sqrt{v}$ [Eq.~\eqref{eq:hot stuff chicken combo}] off and on the horizon, respectively, and the Aretakis instability is present [Eq.~\eqref{eq:hot stuff combo}].
\end{itemize}
Hertz potentials in the dynamical null basis \eqref{eq:ell and n ingoing time-dependent} always decay (Sec.~\ref{sec:EMGrav} above), implying in particular that scalar invariants decay.

\section{Example: Extremal Kerr-Newman-AdS}\label{sec:examples}

Although our main concern has been with identifying universal features, the results of this paper can also be applied to particular perturbation problems.  We now summarize the recipe and give an example of its use.

The first step in determining the critical tail of an extremal spacetime is to take the near-horizon limit by finding coordinates and gauge such that the limit \eqref{eq:limitaceous} exists and gives a metric of the form \eqref{eq:ansatz}.  These coordinates define the ``critical frequency'' by $\omega=0$, where $\omega$ is conjugate to the $t$ coordinate of the limit.  (In general the critical frequency will be non-zero in some original coordinate system in which the metric was written down.)   The limiting metric fixes the free functions in the general near-horizon metric \eqref{eq:ansatz}, which defines an elliptic equation for angular eigenfunctions (Eq.~\eqref{eq:Y eq} for scalars and Eqs.~(2.20) and (2.29) of \cite{Durkee:2010ea} for electromagnetic or gravitational perturbations), whose spectrum of eigenvalues must be computed.  The effective mass and charge are then given by simple formulas [Eqs.~\eqref{eq:psi ads2 eq} and \eqref{eq:psib eq}], from which the exponents $\hp$ of each mode may be computed using \eqref{scaleit}.  Real exponents (supplementary modes) give rise to a  critical tail with power law decay of $v^{-\hp}$ and $t^{-2\hp}$ on and off the horizon, respectively.   (See Sec.~\ref{sec:summary} for a summary of the details.)  If any of the exponents is complex (a principal mode), then a more detailed analysis of the far region is required to determine the critical tail. One must solve the far equations to determine the effective $\adst$ boundary conditions $\mathcal{N}$ of each mode, as described below Eq.~\eqref{eq:up far matching}.  Then, for each principal mode one must check certain conditions.  For modes satisfying Eq.~\eqref{eq:instability condense}, there will be a condensate-type instability.  For modes in case II of \eqref{eq:principal cases} [but not satisfying \eqref{eq:instability condense}], we have not analyzed the behavior.  For modes in cases I or III of \eqref{eq:principal cases}, the critical tail will have the universal $1/\sqrt{v}$ and $1/t$ behavior on and off the horizon, respectively, whose properties are summarized in Sec.~\ref{sec:summary}.

We now illustrate this procedure in the example of four-dimensional extremal Kerr-Newman-AdS (KN-AdS), which may be written in coordinates $(\tilde t, \tilde r,\theta,\tilde \phi)$ as \cite{Carter:1968ks}
\begin{subequations}\label{eq:a nice thing}
\begin{align}\label{eq:KNAdS metric}
d\tilde s^2 & = - \frac{\Delta}{\Sigma}\left(d\tilde t-\frac{a}{\Xi}\sin^2\theta d\tphi\right)^2+\frac{\Sigma}{\Delta} d\tr^2 + \frac{\Sigma}{\Delta_\theta}d\theta^2+\frac{\Delta_\theta}{\Sigma}\sin^2\theta\left(ad\tilde t - \frac{\tilde r^2+a^2}{\Xi}d\tphi\right)^2, \\
\tilde A & = - \frac{q\tr}{\Sigma}\left(d\tilde t - \frac{a \sin^2\theta}{\Xi } d\tphi\right), \label{eq:vec pot BL}
\end{align}
\end{subequations}
where 
\begin{align}\label{eq:KNAdS stuff}
\Delta = (\tr^2+a^2)\left(1+ \frac{\tr^2}{\ell^2}\right) - 2M \tr +q^2, \quad \Delta_\theta = 1- \frac{a^2}{\ell^2} \cos^2\theta,\quad
\Xi &= 1 - \frac{a^2}{\ell^2}, \quad
\Sigma =\tr^2 + a^2 \cos^2 \theta.
\end{align}
 The spacetime characterizes a rotating black hole when $0<a<\ell$.  The event horizon is located at the outermost real root of $\Delta$, denoted $r_+$.  The complete family depends on four parameters $M,a,q,\ell$, which are interpreted in \cite{Caldarelli:1999xj}.  The extremal family, which is our primary interest here, is the parameter set of solutions for which 
\begin{align}
a^2 = \frac{r_+^2}{1-r_+^2/\ell^2}\left(1+ \frac{3 r_+^2}{\ell^2} - \frac{q^2}{r_+^2} \right), \qquad
M = \frac{ r_+} {1-r_+^2/\ell^2}\left(  \left(1+\frac{r_+^2}{\ell^2}\right)^2- \frac{q^2}{\ell^2}\right). \label{eq:mass and spin ext}
\end{align}
In the extremal case, $\Delta$ may be expanded near the horizon as
\begin{equation}\label{eq:Delta near hor}
    \Delta = \frac{(r_+^2+a^2)r_+^2}{r_0^2}(\tilde r/r_+-1)^2 + O(\tilde r/r_+-1)^3,
\end{equation}
where, following the notation in \cite{Hartman:2008pb}, we have introduced
\begin{equation}\label{eq:r0}
r_0^2=\frac{(r_+^2+a^2)(1-r_+^2/\ell^2)}{1-q^2/\ell^2+3r_+^2/\ell^2(2-r_+^2/\ell^2)}.
\end{equation}

\subsection{Near-horizon geometry}
Following \cite{Hartman:2008pb,Compere:2012jk}, we obtain the near horizon spacetime by changing coordinates and gauge by
\begin{align}
    t =\frac{\tilde t}{r_0}, \qquad x=\frac{\tilde r-r_+}{r_0} \qquad  \phi = \tphi -\Omega_H\tilde t, \qquad A = \tilde A +  \Phi_H d\tilde t,
\end{align}
where the horizon frequency $\Omega_H$ and electrical potential $\Phi_H$ are defined by
\begin{align}
   \label{eq:KNAdS Omega}
 \Omega_H = \frac{a\Xi}{r_+^2+a^2}, \qquad \Phi_H = \frac{qr_+}{r_+^2+a^2}.
\end{align}
Letting $x \to 0$ fixing $tx$ (and $\theta$ and $\phi$) gives the near-horizon geometry as
\begin{subequations}\label{eq:a fine thing}
\begin{align}\label{eq:KNAdS NH metric}
ds^2 & = L^2(\theta)d\hat s^2
+ \sigma(\theta) d\theta^2
+%
\gamma(\theta) \left(d\phi+ k \hat A\right)^2, \\
A &=  Q(\theta)(d\phi+k\hat A),\label{eq:KNAdS NH gauge field}
\end{align}
\end{subequations}
where 
\begin{align}\label{eq:KNAdS NH metric functions}
L^2(\theta)=\frac{\Sigma_+}{\delta},\qquad
\sigma(\theta)=\frac{\Sigma_+}{\Delta_\theta},\qquad
\gamma(\theta)= \frac{\Delta_\theta\sin^2 \theta}{\Sigma_+}\frac{a^2}{ \Omega_H^2},\qquad
 Q(\theta) = q\frac{r_+^2-a^2\cos^2\theta}{2\Sigma_+ r_+ \Omega_H}, \qquad
k=\frac{2\Omega_H r_+}{ \delta},
\end{align}
with 
\begin{align}
    \Sigma_+ := r_+^2 +a^2 \cos^2 \theta, \qquad \delta := \frac{r_+^2+a^2}{r_0^2}. 
\end{align}
In Eq.~\eqref{eq:a fine thing}, $d\hat{s}^2$ and $\hat{A}$ are the boundary-adapted $\adst$ metric and gauge \eqref{eq:ads2 stuff}.

Comparing with the general form \eqref{eq:ansatz}, we identify the fiber as the (topological) sphere covered by coordinates $x^\mu=(y^i,\phi^I)=(\theta,\phi)$.  (In this example of a two-dimensional fiber with a single $U(1)$ symmetry $\phi$, we do not need explicit indices $I$ or $i$, and we denote $y=\theta$.)  Eq.~\eqref{eq:KNAdS NH metric functions} fixes all of the free functions in the general ansatz \eqref{eq:ansatz}.  The critical frequency is by definition $\omega=0$ for the notion of frequency $\omega$ conjugate to the near-horizon coordinate $t$ in the near-horizon gauge $A$.  In terms of the original frequency $\tilde{\omega}$ conjugate to the original coordinate $\tilde{t}$ in the original gauge $\tilde{A}$, this becomes
\begin{align}
\textrm{critical frequency: } \qquad \tilde{\omega}= m \Omega_H + e \Phi_H.
\end{align}
In the original coordinates, the critical tail will be associated with time-dependence of this characteristic frequency [Eq.~\eqref{eq:an interesting thing} below].

\subsection{Elliptic equation and exponents}

We consider scalar perturbations for simplicity.  Using Eq.~\eqref{eq:KNAdS NH metric functions}, the elliptic equation \eqref{eq:Y eq} for the near-horizon angular eigenfunctions $Y$ is given by
\begin{equation}\label{eq:fiber eq KNAdS}
    \pd_\theta\left(\Delta_\theta\pd_\theta Y\right)+\Delta_\theta\cot\theta \, 
    \pd_\theta Y - \left( \frac{\Omega_H^2\Sigma_+^2(m-eQ)^2\,\csc^2\theta}{a^2\Delta_\theta}+\mu^2 \Sigma_+  - \mathcal E \delta \,\right) Y =0.
\end{equation}
The spectrum $\mathcal{E}$ can be determined by solving this equation.  In the limit of a massless perturbation to Kerr, $\mathcal E-m^2=K$, where $K$ is 
given in \emph{Mathematica} by $\mathsf{SpheroidalEigenvalue}[l,m,im/2]$.

The effective mass and charge of the $\adst$ perturbations are given by Eq.~\eqref{eq:psi ads2 eq}, 
\begin{equation}\label{eq:KNAdS to AdS2 params}
    \hat \mu^2 = \mathcal E, \qquad \hat e =km=2\Omega_Hr_+m/\delta, 
\end{equation}
and the scaling dimension $\hp$ is [from \eqref{eq:hpm NHG} or \eqref{scaleit}]
\begin{equation}\label{eq:KNAdS h}
    \hpm = \frac{1}{2} + \sqrt{1/4+\mathcal E - (2\Omega_Hr_+m/\delta)^2}.
\end{equation}
In the Kerr-Newman limit, where the $\mathrm{AdS}_4$ length scale $\ell$ is taken to infinity, $\delta=1$ and $\Omega_H$ is given by \eqref{eq:KNAdS Omega} with $\Xi=1$. To further take the RN limit, where $\ell\to\infty$ and $a\to 0$, an additional change of gauge is needed \cite{Hartman:2008pb}.

\subsection{Matching to the full geometry}
In the full KN-AdS geometry, the massive charged scalar equation
\begin{equation}
  \left(\tilde D^2 - \mu^2 \right)\Phi =0, \qquad  \tilde  D = \tilde  \nabla - i e \tilde  A,
\end{equation}
separates under the harmonic decomposition
\begin{equation}
\Phi =   e^{-i\omega \tilde t} \tilde R(\tilde r)\tilde{Y}(\theta,\phi), \qquad \tilde{Y} := e^{ i m \tphi} S(\chi),
\end{equation}
where we have introduced $\chi = a \cos \theta$ following \cite{Chambers:1994ap}.  The $\omega$-dependent functions $S$ satisfy the massive spheroidal equation of the Heun type with eigenvalue $K$
\begin{equation}\label{spheroidal eqn}
  \pd_\chi ( \Delta_\chi \pd_\chi S ) - \left( \frac{ P^2}{\Delta_\chi} +\mu^2 \chi^2 - K \right) S = 0,
\end{equation}  
where $P= \left(\omega+m\Omega_H+e\Phi_H\right) (a^2 - \chi^2) -am(1-a^2/\ell^2)$
and $\Delta_\chi = (1-\chi^2/\ell^2)(a^2 - \chi^2)$.  This equation may be solved using Sturm-Liouville methods by imposing regularity at the poles. 

The radial equation  is
\begin{align}\label{eq:KNAdS radial eq}
\partial_{\tilde r}(\Delta \partial_{\tilde{r}} \tilde R) + \left( \frac{Z^2}{\Delta} -\mu^2 \tr^2 - K  \right) \tilde R = 0,
\end{align}
where 
\begin{align}\label{eq:Z}
Z = \left(\omega+m\Omega_H+e\Phi_H\right) (\tr^2+a^2) - am (1 - a^2/\ell^2) - eQ\,\tr.  
\end{align}
This equation may be used to determine the effective $\adst$ boundary conditions $\mathcal{N}$, given a choice of true boundary conditions at the  $\textrm{AdS}_4$ boundary.  Unfortunately, the equation is of Heun type on an infinite domain, meaning this analysis would have to proceed numerically.  However, we can still check analyticaly that the far solutions properly match to $\adst$ by taking the far limit $\omega \to 0$ and then examining the small-$x$ behavior.  Setting $\omega=0$ in \eqref{eq:KNAdS radial eq} and then expanding near $x=0$ ($\tilde{r}=r_+$) using the Frobenius method, the Frobenius indices are found to be
\begin{align}\label{eq:nupm}
\hpm = \frac{1}{2}\pm\sqrt{\frac{1}{4}+\frac{\mu^2r_+^2+K|_{\omega=0}}{\delta}-\frac{ \left(2mr_+\Omega_H + e\Phi_H(r_+^2 -a^2)/r_+ \right)^2}{\delta^2}}.
\end{align}
This entails behavior of $x^{-\hpm}$ at small $x$ in the far limit, and must match the $\hpm$ of \eqref{eq:KNAdS h}, determined in the near limit.  Comparing the two expressions fixes the near-horizon eigenvalue $\mathcal{E}$ in terms of the $\omega \to 0$ limit of the general eigenvalue $K$ to be
\begin{equation}\label{eq:K and E}
  \mathcal E  =  \frac{K|_{\omega =0} + (r_+\mu)^2}{ \delta  } - \frac{e\Phi_H(r_+^2-a^2)}{r_+^2\delta^2} \left( 4m r_+^2\Omega_H+ e\Phi_H(r_+^2-a^2)\right).
\end{equation}
With this identification, and the previous identifications \eqref{eq:KNAdS to AdS2 params}, the far limit of a generic solution satisfies
\begin{equation}
    \tilde R^{\rm far}  \sim D_+ x^{-\hp} + D_- x^{-\hm}, \quad x\to 0,
\end{equation}
for some constants $D_\pm$.  This means that far solutions will properly match to near region solutions, as assumed in Sec.~\ref{sec:crit freq}.

\subsection{Critical tail}\label{sec:KN-AdS crit tail}

Having verified that a near-far match can be achieved, the next logical step is to determine the full details of the critical tail by computing the eigenvalues $\mathcal{E}$ and the effective near-horizon boundary conditions $\NO$.  This was done for the Kerr spacetime in Ref.~\cite{Gralla:2017lto}.  In this more general setting we do not attempt this calculation and instead  present the range of possible critical tails.  This entails simply utilizing the formulas \eqref{eq:KNAdS h} or \eqref{eq:nupm} for $\hp$ in the general results summarized in Sec.~\ref{sec:summary}.  For example, for a supplementary mode at fixed $\theta$ and $x>0$, the critical tails are
\begin{subequations}
\begin{align}
    \Phi & \propto t^{-2\hp} e^{i m \phi},\qquad \qquad \qquad \qquad \quad\quad\,\textrm{(near-horizon coordinates and gauge \eqref{eq:a fine thing})},\\ 
    & \propto \tilde{t}^{-2 \hp} e^{-i (m \Omega_H+e\Phi_H) \tilde{t} } e^{i m \tilde{\phi}}, \qquad \qquad \textrm{(original coordinates and gauge \eqref{eq:a nice thing})},\label{eq:an interesting thing}
\end{align}
\end{subequations}
 where $k$ was given previously in \eqref{eq:KNAdS NH metric functions}.  Notice the appearance of the phase $e^{- i m \Omega_H \tilde{t}}$ when we re-express in the original coordinates and gauge.  Including also the on-horizon result, we may summarize the situation for supplementary modes as
\begin{align}\label{eq: a lovely thing}
    \Phi \propto \begin{cases} 
    v^{-\hp-ikm}, & \textrm{on a horizon generator in the regular gauge \eqref{eq:a fine thing}} \\
    \tilde{t}^{-2 \hp} e^{-i( m \Omega_H + e \Phi_H) \tilde{t} }, & \textrm{fixed } (\tilde{r}, \theta, \tilde{\phi}) \textrm{ off the horizon in the original gauge \eqref{eq:a nice thing}},
    \end{cases}
\end{align}
where $v$ is an affine parameter on the horizon generators.   

For principal modes (of case I or case III in \eqref{eq:principal cases}), the analysis is similar, using the more complicated phase structure summarized in Sec.~\ref{sec:summary}.  The critical decay is like $1/\tilde{t}$ and $1/\sqrt{v}$ off and on the horizon, respectively.  Since the formula \eqref{eq:nupm} for $\hp$ clearly indicates the potential for principal modes, we can expect to see these universal rates at least in some region of parameter space.  Indeed, $1/\tilde t$ and $1/\sqrt{v}$ decay is known for massless perturbations of extremal Kerr \cite{Gralla:2017lto}, while $1/\tilde t$ intermediate-time behavior was seen for certain massive, charged perturbations of extremal Kerr-Newman \cite{Konoplya:2013rxa}.

\section*{Acknowledgements}
We thank Sean Hartnoll for helpful conversations. This work was supported by NSF grant 1506027 to the University of Arizona.

\appendix

\section{Extremal planar Reissner-Nordstr\"om AdS (RN-AdS)}\label{sec:planar RN-AdS}
RN-AdS black holes have an important place in the AdS/CFT correspondence,
as they are solutions to higher-dimensional supergravity truncations with appropriate compactifications \cite{Romans:1991nq,Chamblin:1999tk}. Here we consider the planar limit of the extremal RN-AdS solution, which plays a role in holographic models for certain condensed matter systems near their quantum critical point \cite{Hartnoll:2016apf}.  Although our framework  has assumed a compact horizon, most of the calculations remain relevant to the non-compact horizon case.  Here we briefly review perturbations of extremal planar RN-AdS black holes to expedite comparison with the holographic condensed matter literature. 

In planar static coordinates $(\tau, r, \vec y)$, the RN-AdS solution is given in $d$ spacetime dimensions by \cite{Chamblin:1999tk}
\begin{subequations}
\begin{align}\label{eq:RN AdS stuff}
    d\tilde s^2&=-Nd\tau^2+N^{-1}dr^2+(r/\ell)^2 d\vec{y}^2, \qquad N=\frac{r^2}{\ell^2}-\frac{m}{r^{d-3}}+\frac{q^2}{r^{2d-6}},\\  \tilde A&=\mathcal{Q}\left(1-\frac{r_+^{d-3}}{r^{d-3}}\right) \mathrm d \tau,
\end{align}
\end{subequations}
where $r_+$ is the location of the outer horizon and
\begin{equation}\label{eq:RN AdS params}
     \mathcal{Q} :=  \frac{q}{r_+^{d-3}}\sqrt{\frac{d-2}{2(d-3)}} \qquad \ell^2 =  -\frac{(d-2)(d-1)}{2 \Lambda}.
\end{equation}
Here $\Lambda$ is the cosmological constant. Using the standard relation for the Hawking temperature $T = N'(r_+)/(4\pi)$ we find that, at extremality $(T=0)$,
\begin{equation}\label{eq:F RNADS}
    N=\frac{r_+^2}{\ell_2^2}(r/r_+-1)^2+O\left(r/r_+-1\right)^3, \qquad \,
\end{equation}
where $\ell_2^2=\ell^2/((d-2)(d-1))$ is the square of the scalar curvature of $\adst$. 

The near-horizon geometry of the extremal solution may be obtained by introducing coordinates 
\begin{equation}
    x= \frac{r-r_+}{\ell_2^2}, \quad t = \tau,
\end{equation}
and taking $x\to 0$ fixing $tx$ as usual. The resulting metric is $\mathbb{R}^{d-2} \times \adst$
\begin{equation}\label{eq:RNADS NHG}
    ds^2= \ell_2^2 d\hat s^2+ \frac{r_+^2}{\ell^2} d\vec y^2,
\end{equation}
where $d\hat s^2$ is given in \eqref{eq:ads2 stuff}. The near-horizon gauge field is related to the $\adst$ potential $\hat A=xdt$ by
\begin{equation}
    A =  Q \hat A, \qquad  Q = \frac{(d-3)\mathcal{Q}\ell^2_2}{r_+}.
\end{equation}

Consider now a charged field $\Phi$ with mass $\mu$ satisfying 
\begin{equation}
   \left( \tilde{D}^2-\mu^2 \right)\Phi=0, \quad \tilde D := \tilde \nabla - i e \tilde A.
\end{equation}
We utilize the planar and time-translation symmetries by adopting the mode-decomposition
\begin{equation}
    \Phi(\tau,r,\vec y) = \int \frac{d^{d-1}k}{(2\pi)^{d-1}}\, \tilde R_{k}(r) \,e^{i \vec k \cdot \vec y-i\omega \tau},
\end{equation}
where the momentum vector $k$ is given by $k =(\omega, \vec k)$ and $\vec k \cdot \vec y$ denotes the Euclidean scalar product on the $d-2$-dimensional transverse space.  This gives rise to the radial equation
\begin{equation}\label{eq:popeye}
   N \tilde R_k'' + \frac{(d-2)N}{r} \tilde R_k' + N'  \tilde R_k'
   + \Bigg[\left(\frac{\omega+e \tilde A_\tau}{\sqrt{N}}\right)^2 - \left(\frac{\ell\, \vec k}{r}\right)^2 -\mu^2 \Bigg]\tilde R_k=0,
\end{equation}
where prime denotes the radial derivative. 
For the zero-temperature (extremal) background, setting $\omega=0$ in \eqref{eq:popeye} results in a radial equation which is difficult to work with. However, by taking the limit $r\to r_+$ 
at $\omega=0$ using \eqref{eq:F RNADS} we see that the solutions must have the asymptotics
\begin{equation}
    \tilde R_{\rm far} \sim D_+ x^{-\hp} + D_- x^{-\hm},
\end{equation}
(for some coefficients $D_\pm$, different for each solution), where 
\begin{equation}\label{eq:Delta pm RNAdS p}
    \hpm = \frac{1}{2}\pm\sqrt{\frac14 + \ell_2^2 \mu^2 + \ell_2^2 (\ell \vec k/r_+)^2 - e^2 Q^2}.
\end{equation}
From Eq.~\eqref{eq:Delta pm RNAdS p} we can read off the effective $\adst$ mass and charge as
\begin{equation}
    \hat\mu^2 = \ell_2^2 \mu^2 + \ell_2^2 (\ell \vec k/r_+)^2, \qquad \hat e^2 = e^2 Q^2.
\end{equation}
The analysis of the critical tail then proceeds identically, giving the behaviors summarized in Sec.~\ref{sec:summary} with the formula \eqref{eq:Delta pm RNAdS p} for $\hp$.  The key difference from the compact case is that the modes are labeled by a continuous parameter $\vec{k}$ instead of a discrete list $\mathcal{E}$.  Correspondingly, each mode does not have compact spatial support.  This makes it more difficult to generalize conclusions about individual modes to generic perturbations.  Work in this direction is underway \cite{GRZsoon}. 

\section{Discrete modes}\label{sec:pain in the ax}

Modes with weight $h_+ + i\hat{e} \in \mathbb{Z}^{>0}$ are called discrete in our classification [Tab.~\ref{tab:h}].  We have excluded these modes from consideration above, and we cannot treat  them as limiting cases, since the coefficient $C_+$ \eqref{eq:Cpm} appearing in the front of the power law tails vanishes in the discrete limit.  Direct treatment of discrete modes in $\adst$ reveals that the mode functions are analytic at $\omega=0$ and hence do not give any power law tail.  The present framework does not make a universal prediction for discrete modes.  

However, discrete modes do in general possess power law tails (and the Aretakis instability), at least in the well-studied cases of four-dimensional, asymptotically flat black holes.  Examples of discrete modes include axisymmetric perturbations of the extremal Kerr spacetime \cite{Casals:2016mel}, neutral (massless) scalar mode perturbations of extremal Reissner-Nordstr\"om \cite{Zimmerman:2016qtn} and axisymmetric gravitational perturbations of higher-dimensional extremal rotating black holes and black rings \cite{Murata:2011my}.  In the four-dimensional cases we have studied in detail, the matched asymptotic expansion for the discrete modes differs from the one we assume here (Sec.~\ref{sec:crit freq}) in two important ways.  First, the far expansion must be defined as $\omega \to 0$ fixing $x \omega$ (instead of $\omega \to 0$ fixing $x$) in order to satisfy the outgoing conditions that define the $\infty$ solution.  Second, in both near and far expansions one must keep subleading terms (in $\omega$) in the equations of motion in order to satisfy all boundary and matching conditions.  The correct near functions are Whittaker functions with an effective frequency-dependent charge $\hat{e}=\omega-ib$, which do not satisfy the wave equation in $\adst$.  Thus the power law tails (and the Aretakis instability) of the discrete modes do not arise from the physics of $\adst$.  It would be interesting to determine whether these modes are controlled by a corrected or otherwise deformed near-horizon geometry. 

\bibliography{MyReferences.bib}
\end{document}